\documentclass[12pt]{iopart}

\def\>{\rangle}
\def\<{\langle}
\def\ket#1{|#1\>}
\def\bra#1{\<#1|}
\def\braket#1#2{\< #1 | #2 \>}
\def\ave#1{\< #1\>}
\def\ve#1{{\bm{#1}}}
\def\ma#1{{\rm \mathbf{#1}}}
\newcommand{\op}[1]{\hat{#1}}

\usepackage{bbm}
\usepackage{bm}
\usepackage{graphicx}

\begin{document}

\title{Stability of quantum motion and correlation decay}

\author{Toma\v z Prosen \footnote{prosen@fiz.uni-lj.si} and Marko \v Znidari\v c \footnote{znidaricm@fiz.uni-lj.si}}

\address{Physics Department, Faculty of Mathematics and Physics, 
University of Ljubljana, Jadranska 19, SI-1000 Ljubljana, Slovenia}


\begin{abstract}
We derive a simple and general relation between the fidelity of quantum motion,
characterizing the stability of quantum dynamics with respect to arbitrary 
static perturbation of the unitary evolution propagator, and the integrated 
time auto-correlation function of the generator of perturbation. 
Surprisingly, this relation predicts the slower decay of fidelity the faster decay of
correlations is. In particular, for non-ergodic and non-mixing dynamics, where
asymptotic decay of correlations is absent, a qualitatively different and faster decay
of fidelity is predicted on a time scale $\propto 1/\delta$ as opposed to
mixing dynamics where the fidelity is found to decay exponentially on a time-scale 
$\propto 1/\delta^2$, where $\delta$ is a strength of perturbation.
A detailed discussion of a semi-classical regime of small effective values of
Planck constant $\hbar$ is given where classical correlation functions can be used to 
predict quantum fidelity decay. Note that the correct and intuitively 
expected classical stability behavior is recovered in the classical limit 
$\hbar\to 0$, as the two limits $\delta\to 0$ and $\hbar\to 0$ do not commute. In 
addition we also discuss non-trivial dependence on the number of degrees of freedom.
All the theoretical results are clearly demonstrated numerically on a celebrated 
example of a quantized kicked top.
\end{abstract}

\noindent
To appear in \JPA (March 2002)

\section{Introduction}

The precise quantum signatures of different qualitative types of classical
motion and the very definition of chaos in quantum mechanics are still the
issues of an unsettled discussion 
(see e.g. \cite{Haake1,Peres1,Nakamura}).
Due to unitarity of quantum dynamics, 
quantum chaos cannot be defined in the same way as the classical chaos 
\cite{casati86}, 
namely through the exponential sensitivity on the variation of initial conditions.
However, Peres \cite{Peres2} proposed an alternative concept which can be used in
classical as well as in quantum mechanics: 
One can study the stability of quantum motion with respect to a small variation in the 
Hamiltonian, or more generally, a variation of the unitary evolution operator.
Clearly, in classical mechanics this concept, when
applied to individual trajectories (or to phase space distribution functions as we show
below in sect.~4), is equivalent to the sensitivity to initial conditions: 
Integrable systems with regular orbits are stable against small variation in the 
hamiltonian (the statement of KAM theorem), wheres for chaotic orbits varying the
hamiltonian has a similar effect as varying the initial condition: 
exponential divergence of two orbits for two nearby chaotic hamiltonians. 

This paper is devoted to a systematic theoretical study of the stability of a
unitary time evolution with respect to small static variations of the unitary 
propagator. It will be primarily applied to the Schr\" odinger propagator 
in quantum dynamics (sect.~2 and 3), however an alternative application to the stability of 
classical unitary Perron-Frobenius evolution will be considered for comparison (sect.~4). 
The quantity of the central interest here is the {\em fidelity} 
of unitary (quantum) motion.
Let us consider a unitary operator $U$ being either (i) a short-time propagator 
$U = \exp(-i H \Delta t/\hbar)$, or (ii) a Floquet map 
$U = \op{\cal T}\exp(-i\int_0^p d\tau H(\tau)/\hbar)$ of 
(periodically time-dependent) Hamiltonian 
$H$ ($H(\tau+p)=H(\tau)$), or (iii) a quantum Poincar\' e map or any other quantized
symplectic map. In any case a general small perturbation of the unitary operator 
can be written in the following
form
\begin{equation}
U_\delta = U\exp(-i A\delta/\hbar), 
\label{eq:U_d}
\end{equation}    
where $A$ is some self-adjoint operator, $\delta$ is a perturbation strength
and $\hbar$ is an effective Planck constant which is taken out explicitly so that
the observable $A$ has a well defined classical limit (e.g. the corresponding Weyl symbol).
The influence of a small perturbation to the unitary evolution
is quantitatively described by the overlap $\braket{\psi_\delta(t)}{\psi(t)}$ measuring 
the Hilbert space distance between the exact and the perturbed time evolution 
from the same initial pure state 
$\ket{\psi(t)} = U^t\ket{\psi}$, 
$\ket{\psi_\delta(t)} = U^t_\delta\ket{\psi}$, 
where {\em integer} $t$ is a
discrete time (in units of the period $p$).
This defines the {\em fidelity} 
\begin{equation}
F(t) = \ave{U_\delta^{-t} U^t},
\label{eq:Ft}
\end{equation}
where $\ave{}$ gives the expectation value in the initial state $\ket{\psi}$.
More generally, it may be useful to statistically average the fidelity over an ensemble
of {\em different pure initial states} $\ket{\psi_k}$ appearing with {\em probabilities} $\rho_k$.
Thus we will write the fidelity in terms of a {\em statistical density operator} 
$\rho = \sum_k \rho_k \ket{\psi_k}\bra{\psi_k}$,
namely as eq.~(\ref{eq:Ft}) using the conventional statistical expectation value 
$\ave{B} = \tr \rho B$.
The theoretical discussion in this paper is fully general, 
however, we will later apply our theory in the two extreme situations,
namely for the (coherent) pure initial state $\rho=\ket{\psi}\bra{\psi}$, 
and for the full Hilbert space average $\rho=\mathbbm{1}/{\cal N}$ which is also
equivalent to considering a {\em random} pure initial state of maximal information entropy. 
Integer ${\cal N}$ denotes a dimension of the relevant Hilbert space which can be 
written semiclassically by the Thomas-Fermi rule
\begin{equation}
{\cal N} = (2\pi\hbar)^{-d}{\cal V}
\end{equation}
where ${\cal V}$ is the classical phase-space (or energy shell) volume and $d$ is the number of
(relevant) degrees of freedom.

The quantity $F(t)$, or its equivalent definitions, have already been discussed in several
different contexts in quantum physics. We name just a few, namely those which provided 
direct motivation for the present work:
First, $F(t)$ has been proposed by Peres \cite{Peres2} 
as a measure of the stability of quantum motion. Second, $|F(t)|^2$ is the {\em Loschmidt echo} 
characterizing the {\em dynamical irreversibility}, which has been used  
e.g. in spin-echo experiments \cite{Usaj} where one is interested in the overlap 
between the initial state $\ket{\psi}$ and an {\em echo} state 
$U_\delta^{-t} U^t\ket{\psi}$ obtained by composing forward time evolution, 
imperfect time inversion with a residual (non-invertible) interaction described by the 
operator $A\delta$, and backward time evolution.
Third, the fidelity has become a standard measure characterizing the loss of phase coherence in 
the quantum computation \cite{qcomp}. Fourth, it was used to characterize the
``hypersensitivity to perturbations'' in related studies of information theoretic 
characterization of quantum chaos \cite{Shack94}, though in a different context of 
a stochastically time-dependent perturbation.

The main result of this paper is a relation of the fidelity to the
ergodic properties of quantum dynamics, more precisely to the time 
autocorrelation function of the generator of perturbation $A$. 
Quantum dynamics of finite and bound systems has always a {\em discrete 
spectrum} since the effective Hilbert space dimension ${\cal N}$ is finite, 
hence it is {\em non-ergodic} and {\em non-mixing} \cite{qmix}: 
time correlation functions have fluctuating tails of order $\sim 1/{\cal N}$.
In order to reach a genuinely complex quantum motion with a continuous spectrum 
one has to enforce ${\cal N}\to\infty$ by considering 
one of the following two limits: semi-classical limit of effective 
Planck's constant $\hbar\to 0$, or thermodynamic limit of number of 
particles, or number of freedoms $d\to\infty$. Our result is surprising 
in the sense that it predicts the fidelity to decay slower if the 
integrated time correlation function is smaller, i.e.
if the relaxation process is faster. 
In particular, for ergodic and sufficiently mixing dynamics (such that the
time integrated autocorrelation function of the perturbation $A$ is finite) 
the fidelity is found to decay {\em exponentially} on a time-scale 
$\propto \delta^{-2}$, whereas for a `more regular', non-ergodic dynamics
with non-vanishing time-averaged correlation functions, the fidelity decay is found
to be qualitatively faster with a characteristic time-scale $\propto\delta^{-1}$.
However, this surprising and apparently counterintuitive result is correctly
reconciled with the expected corresponding behavior in the classical limit
due to non-trivial non-interchangability of the limits $\delta\to 0$ and
$\hbar\to 0$. In all cases, the relevant time-scale for the decay of fidelity
in the semi-classical regime can be explicitly computed in terms of the classical 
quantities only (e.g. classical correlation functions).
The main theoretical predictions are clearly demonstrated on the
numerical examples of a kicked top and a pair of coupled kicked tops.
Furthermore, our theory on fidelity is general and can be applied 
to any perturbed unitary evolution; as an example we consider the 
stability of the classical unitary Perron-Frobenius (Liouville) evolution for maps, 
where we show how `classical fidelity' behaves in a qualitatively
different way than the quantum fidelity and conforms to the
corresponding linearly stable and exponentially unstable behaviors in the 
respective limiting cases of regular and chaotic dynamics.
Two short announcements of our theory with applications in the contexts of
many-body quantum dynamics and quantum computing have already been reported
\cite{Prosen01,QC}.

In section 2, we present a theoretical derivation of the general relation between 
the fidelity of unitary motion and the correlation decay, discuss qualitatively 
different regimes of stability in the semiclassical range of small effective $\hbar$ 
and outline the corresponding time-scales. In section 3, our theoretical predictions are 
demonstrated in detail by numerical experiments on a quantized kicked top. 
In addition, we consider also a pair of coupled kicked tops in order to illustrate some 
dimensionality dependent aspects. In section 4, we demonstrate the conceptual difference 
between the stability of classical and quantum unitary evolution by applying our theory 
to the case of classical (Liouvillian) unitary evolution of phase space densities. 
In section 5, we conclude by pointing to some potentially important 
applications and implications of our results.

\section{General theory}

We start by rewriting the fidelity (\ref{eq:Ft}) in terms of the 
Heisenberg evolution of the perturbation $A_t:=U^{-t}AU^t$. 
Let us recursively insert the unity $U^{t'}U^{-t'}$ in the definition (\ref{eq:Ft}) for $F(t)$ 
and observe $U^{-t'}U_\delta^{-1}U^{1+t'} = \exp(i A_{t'}\delta/\hbar)$
for $t'$ running from $t-1$ downto $0$, yielding
\begin{eqnarray}
F(t) &=& \ave{U_\delta^{-t} U^t} = \ave{U_\delta^{-(t-1)} 
U^{t-1}U^{-(t-1)}(U_\delta^{-1} U) U^{t-1}} = \nonumber \\
 &=& \ave{U_\delta^{-(t-1)}U^{t-1}\exp(i A_{t-1}\delta/\hbar)} = \nonumber \\
 &=& \ave{U_\delta^{-(t-2)}U^{t-2}\exp(i A_{t-2}\delta/\hbar)\exp(i A_{t-1}\delta/\hbar)} = \nonumber \\
 &\ldots&\nonumber \\
 &=& 
\ave{\exp(i A_0\delta/\hbar)\exp(i A_1\delta/\hbar)\cdots\exp(i A_{t-1}\delta/\hbar)}. 
\label{eq:Fprod}
\end{eqnarray}
The obvious next step is to expand the product (\ref{eq:Fprod}) into a power-series
in $\delta$
\begin{equation}
F(t) = 1 + \sum_{m=1}^\infty \frac{i^m \delta^m}{m!\hbar^m}
 \op{\cal T} \!\!\! \sum_{t_1,\ldots,t_m=0}^{t-1} \ave{ A_{t_1} A_{t_2} \cdots A_{t_m} },
\label{eq:Fsum}
\end{equation}
where the operator $\op{\cal T}$ denotes a left-to-right time ordering. 
Such a perturbative expansion converges absolutely for any $\delta$ provided only that the perturbation 
$A$ is a bounded operator. Therefore, the fidelity may be computed by an arbitrary 
truncation of the above expansion (\ref{eq:Fsum}) where its order $M$ ($m\le M$) depends only on the
desired accuracy ${\cal O}(\delta^{M+1})$. We can see that the fidelity $F(t)$ can be expressed entirely in terms of 
time correlation functions of the generator $A$ of perturbation. 

Before proceeding further we note that the first order term ($m=1$) can be trivially eliminated, namely 
by transforming the perturbations as $A\to A-a\mathbbm{1}$ with $a:=(1/t)\sum_{t'=0}^{t-1}\ave{A_{t'}}$ 
(i.e. the perturbation $A$ has to be {\em traceless} if $\rho=\mathbbm{1}/{\cal N}$). 
This does not affect the absolute value $|F(t)|$ since a shift by a multiple
of unity reflects in a simple complex rotation of fidelity $F(t)\to F(t)\exp(-i\delta a t/\hbar)$.
So, to the lowest non-trivial order ($M=2$) we find
\begin{equation}
F(t)=1-\frac{\delta^2}{2\hbar^2} \sum_{t',t''=0}^{t-1} \op{\cal T}C(t',t'') + {\cal O}(\delta^3),
\label{eq:F2nd}
\end{equation}  
where 
\begin{equation}
C(t',t'') := \ave{A_{t'} A_{t''}},{\qquad} \op{\cal T}C(t',t'')=C({\rm min}\{t',t''\},{\rm max}\{t',t''\}),
\end{equation}
is a 2-point time correlation function of the quantum observable $A$. 
In the following, the regime of sufficiently small $\delta$, such
that the fidelity remains close to $1$ (i.e. $1-F(t)\ll 1$) and the above formula (\ref{eq:F2nd}) is 
accurate, will be referred to as the regime of {\em linear response}. 
Although we have a general convergent formula for systematic higher order corrections (\ref{eq:Fsum}) 
we note that all the qualitative and in most cases even quantitative results of this paper can be explained by 
this simple linear-response Kubo-like formula (\ref{eq:F2nd}).

An interesting and somewhat counterintuitive conclusion can be drawn from
the expression (\ref{eq:F2nd}), namely the {\em smaller} the time-integrated
correlation function the {\em higher} the fidelity.
In the semiclassical regime of approaching the classical limit $\hbar\to 0$ 
the quantum correlation function $C(t',t'')$ goes over to the corresponding classical correlation 
function provided a 
state $\rho$ converges to some classical phase space distribution, like e.g. for $\rho=\mathbbm{1}/{\cal N}$.

Therefore, the fidelity for classically chaotic systems will decay with the rate which is
{\em inversely proportional} to their rate of mixing, and furthermore
for classically non-ergodic, i.e. regular or integrable motion, the correlation functions will 
generally not decay to zero and the fidelity will therefore decay much faster. 
Of course, this surprising general conclusions can only be true if the initial state $\ket{\psi}$
(or the statistical density matrix $\rho$) is {\em sufficiently random}, i.e. if it is not a small
linear combination of the eigenstates of $U$ (or $U_\delta$) as will be discussed in more detail later.
Now let us have a closer look at fidelity in the two qualitatively different regimes of correlation decay.

\subsection{Regime of ergodicity and fast mixing}

Here we assume that the system is (classically) ergodic and mixing such that the
correlation function of the perturbation $A$ decays sufficiently fast; this typically 
(but not neccessarily!) corresponds to globally chaotic classical motion.
Due to {\em ergodicity} we will safely assume $\rho=\mathbbm{1}/{\cal N}$,
i.e. average over the full Hilbert space\footnote{In case of autonomous (time-independent) systems,
canonical or micro-canonial state should be used instead.} $\ave{.}=\tr(.)/{\cal N}$. 
For any other initial state (e.g. in the worst case for the minimal wavepacket -- coherent state) 
one obtains identical results on $F(t)$ for sufficiently long times, i.e.
longer than the {\em Ehrenfest -- ergodic} time $t_{\rm E} \approx \ln(1/\hbar)/\lambda$ (for a
classically chaotic system with maximal Lyapunov exponent $\lambda$) needed for a minimal wavepacket
to spread effectively over the accessible phase space.
The state averaged quantum correlation function is homogeneous in time $C(t',t'')=C(t''-t')$ so
we simplify the second order -- linear response -- formula for the fidelity
\begin{equation}
F(t)=1-\frac{\delta^2}{\hbar^2}\left\{\frac{1}{2}tC(0) + \sum_{t'=1}^{t-1}{(t-t')C(t')} \right\} + {\cal O}(\delta^3),
\label{eq:FC}
\end{equation}
and note again that we have assumed $A$ to be traceless $\ave{A}=0$.

If the decay of correlation function $C(t)$ is sufficiently fast, namely it is sufficient that the sums
$I_k=\sum_{t=-\infty}^\infty |t^k C(t)|$ exist for $k\in\{0,1\}$
and that a certain characteristic {\em mixing time} exists, e.g. 
$t_{\rm mix}=I_1/I_0$, then the above formula can be further simplified. 
For times $t \gg t_{\rm mix}$ we can neglect the second term under the summation in (\ref{eq:FC}) and obtain a
linear decay in time $t$ in the linear response regime
\begin{equation}
F_{\rm em}(t) = 1 - (\delta/\hbar)^2 \sigma t
\qquad{\rm with}\qquad
\sigma = \frac{1}{2}C(0) + \sum_{t=1}^\infty{C(t)}
\label{eq:sigma}
\end{equation}
which is a transport coefficient given by the integral of correlation function.
We can make a stronger statement in a non-linear-response regime if we make an additional assumption on the
factorization of higher order time-correlations -- $n-$point mixing. This implies that $2m$-point correlation
$\ave{A_{t_1}\cdots A_{t_{2m}}}$ is appreciably different from zero for $t_{2m}-t_1 \to \infty$ only if all the 
(ordered) time indices $\{t_j,j=1\ldots 2m\}$ are {\em paired} with the time differences 
within each pair $t_{2j}-t_{2j-1}$ being of the order or less than 
$t_{\rm mix}$. Then we can make a further reduction, namely if $t\gg m t_{\rm mix}$ 
\begin{equation}
\!\!\!\!\!\!\!\!\!\!\!\!\!\!\!\!\!\!\!\!\!\!\!\!\!\!\!\!\!
\op{\cal T}\!\!\!\sum_{t_1,\ldots,t_{2m}=0}^{t-1}\!
\ave{A_{t_1} A_{t_2} \cdots A_{t_{2m}}}
\rightarrow
\op{\cal T}\!\!\!\sum_{t_1,\ldots,t_{2m}=0}^{t-1}\!
\ave{A_{t_1} A_{t_2}}\cdots\ave{A_{t_{2m-1}} A_{t_{2m}}}
\rightarrow
\frac{(2m)!}{m!}(t\sigma)^m
\label{eq:factor}
\end{equation}
so that we obtain a global exponential decay 
\begin{equation}
F_{\rm em}(t)=\exp{(-t/\tau_{\rm em})}, \qquad \qquad \tau_{\rm em}=\frac{\hbar^2}{\delta^2 \sigma},
\label{eq:Fmix}
\end{equation}
with a time-scale $\tau_{\rm em}={\cal O}(\delta^{-2})$. 
We should stress again that the above result (\ref{eq:Fmix}) has been derived by the assumption of
true quantum mixing which can be justified only in the limit ${\cal N}\rightarrow\infty$, e.g.
either in semiclassical or thermodynamic limt. However, precise study of finite size effects
and the relevant time and perturbation scales will be given in subsection \ref{sec:timepert}.

\subsection{Non-mixing and non-ergodic regime}

The opposite situation of a non-mixing and non-ergodic quantum dynamics, which typically corresponds to integrable, 
quasi-integrable (KAM), or mixed classical dynamics, is characterized by a non-vanishing time-average of the 
correlation function
\begin{equation}
\bar{C}=\lim_{t\to \infty}{\frac{1}{t^2}\sum_{t',t''=0}^{t-1} \op{\cal T}C(t',t'')}.
\label{eq:Cinfty}
\end{equation}
Here, due to non-ergodicity, time-average $\bar{C}$ depends on the choice of the initial state 
$\ket{\psi}$ or, more generally, on the density matrix $\rho$.
We assume that a characteristic averaging time-scale $t_{\rm ave}$ exists, namely it is an effective time 
$t=t_{\rm ave}$ at which the limiting process (\ref{eq:Cinfty}) converges.
Therefore, for sufficiently large times $t \gg t_{\rm ave}$ the linear-response formula (\ref{eq:F2nd}) 
gives, in contrast to (\ref{eq:sigma}), a simple {\em quadratic decay} in time
\begin{equation}
F_{\rm ne}(t)=
1-\frac{1}{2} \left( \frac{t}{\tau_{\rm ne}}\right)^2 + {\cal O}(\delta^3), \qquad
\tau_{\rm ne}=\frac{\hbar}{\delta \sqrt{\bar{C}}},
\label{eq:Fr2}
\end{equation}
with time-scale $\tau_{\rm ne}={\cal O}(\delta^{-1})$. One should observe that the non-ergodic time-scale 
$\tau_{\rm ne}$ can be much smaller than the ergodic-mixing time-scale $\tau_{\rm em}$ (\ref{eq:Fmix})
provided $\hbar$ is fixed, or the limit $\delta\to 0$ is taken prior to the limit $\hbar\to 0$.
Again, we can make much stronger general statement using an additional assumption. 
Namely, if we assume that $t\gg t_{\rm ave}$ then we can re-write the $m$-tuple sums in the series
(\ref{eq:Fsum}) in terms of a {\em time average perturbation operator}
\begin{equation}
\bar{A}=\lim_{t \to \infty}{(1/t)\sum_{t'=0}^{t-1}{A_{t'}}},
\end{equation}
namely
\begin{equation}
F_{\rm ne}(t)=\sum_{m=0}^\infty {\frac{(i \delta t)^m}{\hbar^m m!} \ave{\bar{A}^m}}
=\ave{\exp{(i t \bar{A} \delta/\hbar)}}.
\label{eq:Favg}
\end{equation}  
Note that $\bar{A}$ is by construction an {\em integral of motion} \cite{PeresInt}, 
$[U,\bar{A}]\equiv 0$, and reduces to a trivial multiple of identity (in fact, it
vanishes with our choice of $\ave{A}=0$) in the ergodic case of previous subsection. Whereas in an ergodic and mixing case, 
$m-$th order term of (\ref{eq:Fsum}) grows with time only as 
${\cal O}(t^{m/2})$ (for even $m$) since it is dominated by pair time correlations, here in a non-ergodic case, the non-trivial time average operator 
$\bar{A}$ already gives the 
dominant effect, namely ${\cal O}(t^m)$ for $m-$th order term of (\ref{eq:Fsum}), so the effect of pair 
time correlations can safely be neglected for sufficiently long times ($t\gg t_{\rm ave}$).
Observe also that time averaged correlation is just a second moment of the time average operator
\begin{equation}
\bar{C} = \ave{\bar{A}^2}.
\end{equation}
In the semiclassical regime, $\bar{C}$ goes to a purely classical, 
$\hbar$-independent quantity 
$\bar{C}_{\rm cl}$ {\em only} for such initial states $\ket{\psi}$,
corresponding to $\hbar$-independent distribution of initial conditions in 
the classical phase space, e.g. for random states or 
$\rho=\mathbbm{1}/{\cal N}$.
Formula (\ref{eq:Favg}) expresses the fidelity in terms of a sequence of moments $\ave{\bar{A}^m}$.
It has been shown \cite{Prosen01} that for a general class of one-dimensional spin chains, the moments
in {\em thermodynamic limit} $d\to\infty$ of long chains tend to a {\em normal} behavior 
$\ave{\bar{A}^{2m}}\to (2m-1)!!\bar{C}^m$, $\ave{\bar{A}^{2m+1}}\to 0$, so fidelity exhibits a
gaussian decay
\begin{equation}
F_{\rm normal}(t) = \exp{(-(t/\tau_{\rm ne})^2/2)}.
\label{eq:Fgauss}
\end{equation}    
Although good numerical agreement with this formula has been observed in several related contexts, e.g. in 
quantum computing \cite{QC}, it cannot be generally valid, in particular not in the semiclassical
regime with a small number of freedoms $d$, as we show below.
\par
In general, we can rewrite the formula (\ref{eq:Favg}) in terms of (average) {\em local density of states} (LDOS)
of the time average $\bar{A}$ which is defined (for finite ${\cal N}$) as
\begin{equation}
d_\rho(a) = \sum_n \delta(a-a_n) \bra{a_n}\rho\ket{a_n}.
\end{equation}
where $a_n$ are the eigenvalues and $\ket{a_n}$ the eigenstates of self-adjont operator $\bar{A}$.
Namely
\begin{equation}
\ave{\exp(i\delta\bar{A}t/\hbar)} = \int\!d a\, e^{i a t\delta/\hbar}d_\rho(a),
\end{equation}
so the fidelity $F_{\rm ne}(t)$ of non-ergodic systems is just a Fourier transformation of LDOS 
$d_\rho(a)$ of the time average perturbation $\bar{A}$ (which becomes the simple (global) density of 
states in case of an uniform average $\rho=\mathbbm{1}/{\cal N}$). 
Thus only a gaussian LDOS would yield a gaussian fidelity decay (\ref{eq:Fgauss}),
however we often find good finite time gaussian behavior which goes beyond the
second moment $\int\!d a\, a^2 d_\rho(a) = \ave{\bar{A}^2} = \bar{C}$,
even if the asymptotic time dependence is completely different,
see e.g. fig.~\ref{fig:analit} for an example of kicked top.
\par
Let us now try to give some semi-classical estimates on the fidelity decay in non-ergodic regime 
by expressing the asymptotics $\hbar\to 0$ in terms of classical information only. Here we assume that our (classical) map 
is integrable so that there exist $d$ canonical constants of motion -- the action variables $\ve{I}=(I_1,\ldots,I_d)$, which
are quantized using EBK rule $\ve{I}_\ve{n}=\hbar(\ve{n}+\ve{\gamma}/4)$ where integers 
$n_j\ge 0,j\in\{1,\ldots,d\}$ are {\em quantum numbers} and integers $\gamma_j$ Maslov indices. 
Since the time average operator $\bar{A}$ commutes with $U$ and with 
the actions $\ve{I}$, it is diagonal in (generically non-degenerate) basis of
eigenstates of $\ve{I}$ (quantized tori) $\ket{\ve{n}}$, namely in the leading semiclassical order
\begin{equation}
\bra{\ve{n}}\bar{A}\ket{\ve{n}'} = 
\delta_{\ve{n},\ve{n}'}\bar{a}(\ve{I}_\ve{n})
\end{equation}
where $\bar{a}(\ve{I})$ is the corresponding classical time-averaged observable in action space.
The fidelity (\ref{eq:Favg}) can therefore be written as
\begin{equation}
F_{\rm ne}(t) = \sum_{\ve{n}}\exp(i t \bar{a}(\ve{I}_\ve{n}) \delta/\hbar)\bra{\ve{n}}\rho\ket{\ve{n}}
\label{eq:Fscs}
\end{equation}
Now, provided the diagonal elements of the density matrix can be written in terms of some smooth function
$ D_\rho(\ve{I}_{\ve{n}}) = \bra{\ve{n}}\rho\ket{\ve{n}}$, and replacing the sum (\ref{eq:Fscs}) by an integral over 
the action space, which is justified for small $\hbar$ up to classically long time $\propto \hbar^0 \delta^{-1}$, 
we obtain
\begin{equation}
F_{\rm ne}(t)=\hbar^{-d} \int\! d^d\ve{I}\, \exp{\{i t \bar{a}(\ve{I})\delta/\hbar\}}D_\rho(\ve{I}) 
\label{eq:Fsemi}
\end{equation}
The obvious next step is to compute this integral by a method of stationary phase.
However, the result depends on the precise form of the function $D_\rho(\ve{I})$ (which may also explicitly 
depend on $\hbar$) so we only work out the details for two important special cases. 

\subsubsection{Semiclassical asymptotics for an uniform initial state average.}

Let us first assume uniform averaging over (random) initial states $\rho=\mathbbm{1}/{\cal N}$, so 
$D_\rho(\ve{I}) \equiv 1/{\cal N} = (2\pi\hbar)^d/{\cal V}$. Then, for large $t \delta/\hbar$ the above integral (\ref{eq:Fsemi})
can be written as a sum of contributions stemming from, say $p$ points, $\ve{I}_\eta, \eta=1,\ldots,p$ where the phase is stationary,
$\partial\bar{a}(\ve{I}_\eta)/\partial\ve{I}=0$,
yielding
\begin{equation}
F_{\rm ne}^{\rm ave}(t) = \frac{(2\pi)^{3d/2}}{\cal V}\left|\frac{\hbar}{t\delta}\right|^{d/2}\sum_{\eta=1}^p
\frac{\exp\{i t\bar{a}(\ve{I}_\eta)\delta/\hbar + i \nu_\eta \}}{|\det \ma{\bar{A}}_\eta|^{1/2}},
\label{eq:Fsqrt}
\end{equation}
where $\{\ma{\bar{A}}_\eta\}_{jk} := \partial^2 \bar{a}(\ve{I}_\eta)/\partial I_j\partial I_k$ is a matrix of second
derivatives at the stationary point $\eta$, and $\nu_\eta = \pi(m_+ - m_-)/4$
where $m_{\pm}$ are the numbers of positive/negative eigenvalues of the matrix
$\ma{\bar{A}}_\eta$. 
We repeat that the stationary phase formula (\ref{eq:Fsqrt}) is expected to be correct in the range 
${\rm const}\,\hbar/\delta < t < {\rm const}'/\delta$.
Most interesting to note is the asymptotic power-law time and perturbation dependence 
$F^{\rm ave}_{\rm ne} \sim |\hbar/(t\delta)|^{d/2}$, which allows for a possible crossover to a 
gaussian decay (\ref{eq:Fgauss}) when approaching the thermodynamic limit $d\to\infty$.

\subsubsection{Semiclassical asymptotics for a coherent initial state.}

Now let us consider a single $d$-dimensional general coherent state centered at $(\ve{I}^*,\ve{\theta}^*)$ in 
action-angle space,
$\rho=\ket{\ve{I}^*,\ve{\theta}^*}\bra{\ve{I}^*,\ve{\theta}^*}$,
\begin{equation}
\!\!\!\!\!\!\!\!\!\!\!\!\!\!\!\!\!\!
\braket{\ve{n}}{{\ve{I}^*,\ve{\theta}^*}} = 
\left(\frac{\hbar}{\pi}\right)^{d/4}
\!\!\!\left|\det\Lambda\right|^{1/4}
\exp\left\{-\frac{1}{2\hbar}(\ve{I}_{\ve{n}} - \ve{I}^*)\cdot\Lambda(\ve{I}_{\ve{n}}-\ve{I}^*) - 
i\ve{n}\cdot\ve{\theta}^*\right\},
\label{eq:CS}
\end{equation}
where $\Lambda$ is a positive symmetric $d\times d$ matrix of squeezing 
parameters, giving 
$D_\rho(\ve{I}) = (\hbar/\pi)^{d/2}\left|\det\Lambda\right|^{1/2}
\exp(-(\ve{I}-\ve{I}^*)\cdot\Lambda (\ve{I}-\ve{I}^*)/\hbar)$ and
\begin{equation}
F^{\rm coh}_{\rm ne}(t) = \frac{\left|\det\Lambda\right|^{1/2}}{(\pi\hbar)^{d/2}}
\int\!d^d\ve{I}\,\exp\left\{\frac{1}{\hbar}
\left(i t \bar{a}(\ve{I})\delta 
- (\ve{I}-\ve{I}^*)\cdot\Lambda(\ve{I}-\ve{I}^*)
\right)\right\}.
\label{eq:Fcohst}
\end{equation}
Using the assumption $\delta t\ll 1$, we see that a unique stationary point $\ve{I}_s$ of 
the exponent approaches $\ve{I}^*$ as $\delta\to 0$, 
\begin{equation}
\ve{I}_s = \ve{I}^* - \frac{it\delta}{2}\Lambda^{-1}\ve{a}' + {\cal O}(\delta^{2}),
\quad {\rm where}\quad
\ve{a}' := \frac{\partial\bar{a}(\ve{I}^*)}{\partial\ve{I}} 
\end{equation}
so we may explicitly evaluate (\ref{eq:Fcohst}) by the method of stationary phase without any lower bound on the range of time $t$,
\begin{equation}
F^{\rm coh}_{\rm ne}(t) = \exp\left\{-\frac{(\ve{a}'\cdot \Lambda^{-1}\ve{a}')\delta^2}{4\hbar}t^2 + 
\frac{i\bar{a}(\ve{I}^*)\delta}{\hbar} t\right\}.
\label{eq:Fcoh}
\end{equation}
Note that the fidelity decay for a coherent initial state with regular classical motion has a time-scale
\begin{equation}
\tau_{\rm ne-coh} = (2\hbar)^{1/2}(\ve{a}'\cdot\Lambda^{-1}\ve{a}')^{-1/2}\delta^{-1}
\;\;\propto\;\; \hbar^{1/2}\delta^{-1}
\label{eq:taunecoh}
\end{equation}
which is consistent with (\ref{eq:Fr2}) with 
$\bar{C}=\frac{1}{2}\hbar(\ve{a}'\cdot\Lambda^{-1}\ve{a}')$
and is typically (for small $\hbar$) larger than the time-scale 
$\tau_{\rm ne}\propto \hbar/\delta$ for a random initial state.
However, surprisingly enough, for sufficiently small $\delta$, fidelity decay for coherent state with regular classical dynamics
can still be faster than the fidelity decay for ergodic and mixing dynamics.
It should be noted that the above derivation for fidelity decay of a coherent state 
(\ref{eq:CS}-\ref{eq:taunecoh}) remains valid in a quasi-integrable (KAM) situation 
of mixed classical phase space, provided that the initial wave packet is launched in a 
{\em regular region} of phase space where (local) action-angle variables exist.

From the derivation of gaussian behavior (\ref{eq:Fcoh}) we can also begin to understand the reason 
for a faster {\em gaussian} decay of fidelity for a non-mixing system in contrast to the 
{\em exponential} decay for a mixing case. 
Namely, in the non-mixing case, the displacement between
centers of the wave packets of two nearby regular evolutions grows
{\em ballistically} and thus produces destructive interference (i.e. decay
of fidelity) much faster than in the case of mixing dynamics where the
destructive interference builds up in a {\em diffusive}, random way.

\subsection{Finite size effects and time and perturbation scales}   

\label{sec:timepert}

The theoretical relations of the previous subsections are strictly justified in the asymptotics ${\cal N}\to\infty$.
However, in any experimentally or practically relevant situation the dimensionality of the relevant Hilbert (sub)space
${\cal N}$ is finite, so we have to discuss in detail the limitations and additional time and perturbation scales 
associated with this.

\begin{figure}
\centerline{\includegraphics{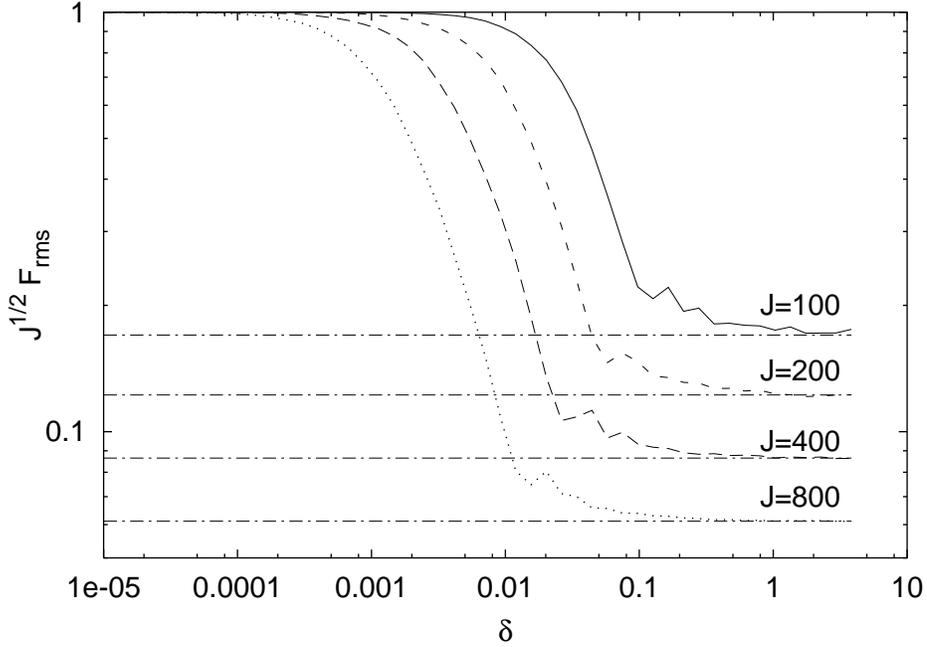}}
\caption{Finite-size fidelity fluctuation for a kicked top 
(\ref{eq:Ukt}) in the mixing regime ($\alpha=30$) and with the whole Hilbert space average (\ref{eq:Frmsc}) is shown
as a function of the perturbation strength $\delta$ at four different values of $J(=1/\hbar)=100,200,400,800$ 
(curves from right to left, respectively). Data are scaled in such a way 
$J^{1/2} F_{\rm rms}$ (note that ${\cal N}=J$) that the weak perturbation regime corresponds to value 
$1$ while the horizontal lines give the corresponding values in the strong perturbation regime 
$F_{\rm rms-iii}^{\rm strong}=\sqrt{3}/J$. Note that the crossover perturbation scales as 
$\delta_{\rm rm}\propto 1/J$.}
\label{fig:Frms}
\end{figure}

First of all, for finite ${\cal N}$, fidelity $F(t)$ cannot decay indefinitely but starts to fluctuate for long times
due to discreteness of the spectrum of the evolution operator $U$. Let us write the eigenphases of $U$ and $U_\delta$, and the
corresponding eigenvectors, respectively, as $\phi_n$, $\phi^\delta_n$, and 
$\ket{\phi_n}$, $\ket{\phi^\delta_n}$, $n=1,\ldots,{\cal N}$, 
satisfying $U\ket{\phi_n}=e^{-i\phi_n}\ket{\phi_n}$,
$U_\delta\ket{\phi^\delta_n}=e^{-i\phi^\delta_n}\ket{\phi^\delta_n}$.
Let us define a unitary operator $V$ which maps the eigenbasis of 
$U$ to the eigenbasis of $U_\delta$, namely $V\ket{\phi_n}:=\ket{\phi^\delta_n}$ for all $n$, with matrix elements
$V_{mn} := \bra{\phi_m}V\ket{\phi_n}$, and write the matrix elements of the statistical averaging operator 
as $\rho_{mn} :=\bra{\phi_m}\rho\ket{\phi_n}$. Note that the matrix $V_{mn}$ is {\rm real orthogonal}, 
$V \in O({\cal N})$,
if $U$ and $U_\delta$ possess a common 
anti-unitary symmetry (e.g. time-reversal).
Now it is straightforward to rewrite the fidelity 
(\ref{eq:Ft}) as
\begin{equation}
F(t) = \sum_{n,l} (\rho V)_{nl} V^*_{nl} \exp\left(i(\phi_n-\phi^\delta_l)t\right),
\label{eq:Ftfin}
\end{equation}
where $(\rho V)_{nl}=\sum_m\rho_{nm} V_{ml}$ is a matrix element of the operator product $\rho V$.
At this point we are interested in the long time fluctuations so we compute the time average 
fidelity fluctuation
\begin{equation}
F_{\rm rms}^2 := \overline{|F|^2} = \sum_{n,l} |(\rho V)_{nl}|^2 |V_{nl}|^2.
\label{eq:Frms}
\end{equation}
In the process of averaging over the time we have assumed that the eigenphases are non-degenerate so
$\overline{\exp(i(\phi_n-\phi_{n'}+\phi^\delta_l-\phi^\delta_{l'})t)} = \delta_{n n'}\delta_{l,l'}$.
So we see that fidelity fluctuation $F_{\rm rms}$ depends on the orthogonal/unitary matrix $V_{mn}$ and the statistical matrix
$\rho_{mn}$. Clearly, $V_{mn}$ converges to an identity matrix for sufficiently {\em small} $\delta$, while for {\em large}
enough $\delta$ and sufficiently complex perturbation $A$ one may assume that it becomes a random orthogonal/unitary matrix.
Of course, the critical perturbation strength $\delta=\delta_{\rm rm}$ for a transition between the two regimes is very much 
system dependent and cannot be discussed in general.
Let us now discuss three different limiting cases of the statistical matrix $\rho_{mn}$:
\\\\
{\bf (i)} First, consider the simplest case where the initial state is a pure eigenstate of $U$ say, 
$\rho=\ket{\phi_1}\bra{\phi_1}$. Then $\rho_{mn} = \delta_{m,1}\delta_{n,1}$ and (\ref{eq:Frms}) rewrites to
\begin{equation}
F_{\rm rms-i}^2 = \sum_l |V_{1l}|^4.
\label{eq:Frmsa}
\end{equation}
Now, for {\em weak perturbation} $\delta\ll\delta_{\rm rm}$ fidelity does not decay at all $F^{\rm weak}_{\rm rms-i} = 1$,
while for {\em strong perturbation} $\delta\gg\delta_{\rm rm}$ we may estimate (\ref{eq:Frmsa}) using the fact that $V_{mn}$ become
real/complex gaussian random variables with variance $1/{\cal N}$ for the orthogonal($\beta=1$)/unitary($\beta=2$) case,
namely $F^{\rm strong}_{\rm rms-i} = [(4-\beta)/{\cal N}]^{1/2}$.
This result can be generalized to the case where initial state can be written as a finite superposition of a small number, say $K$, 
of eigenstates of $U$: then for a weak perturbation $F^{\rm weak}_{\rm rms} \sim K^{-1/2}$. With this simple observation
we can easily explain the numerical result of Peres \cite{Peres2} where no-decay of fidelity was found for a coherent initial state
sitting in the center of elliptic island, thus being a superposition of a very small number of eigenstates (it is almost an eigenstate), 
while the behavior in generic cases may be drastically different as described in the present work.
\\\\
{\bf (ii)} Second, consider the case of a {\em random pure} initial state $\ket{\psi}=\sum_n c_n\ket{\phi_n}$, giving $\rho_{mn}=c_m c_n^*$. 
The principle of
maximal entropy forces the coefficients $c_n$ to be independent random complex gaussian variables with variance $1/{\cal N}$.
Thus, averaging (\ref{eq:Frms}) over $c_n$'s yields
\begin{equation}
F_{\rm rms-ii}^2 = \frac{1}{{\cal N}^2}\left(\sum_{n,l} |V_{nl}|^4 + \sum_{n,m,l}|V_{nl}|^2 |V_{ml}|^2\right).
\label{eq:Frmsb}
\end{equation}
Here, for weak perturbation $\delta\ll\delta_{\rm rm}$ we find $F^{\rm weak}_{\rm rms-ii} = (2/{\cal N})^{1/2}$
while for strong perturbation $\delta\gg\delta_{\rm rm}$ we have 
$F^{\rm strong}_{\rm rms-ii}= {\cal N}^{-1/2}$
\\\\
{\bf (iii)} Third, in the limiting case of a uniform average over all initial states $\rho=\mathbbm{1}/{\cal N}$, 
$\rho_{mn}=\delta_{mn}/{\cal N}$, we have
\begin{equation}
F_{\rm rms-iii}^2 = \frac{1}{{\cal N}^2}\sum_{n,l} |V_{nl}|^4.
\label{eq:Frmsc}
\end{equation}
Again, for weak perturbation $\delta\ll\delta_{\rm rm}$, $F^{\rm weak}_{\rm rms-iii} = {\cal N}^{-1/2}$, and for 
strong perturbation $\delta\gg\delta_{\rm rm}$, $F^{\rm strong}_{\rm rms-iii} = (4-\beta)^{1/2}/{\cal N}$.
\\\\
Note that the formulae (\ref{eq:Frmsa}-\ref{eq:Frmsc}) state simply that fidelity fluctuation is 
an {\em inverse participation ratio} (IPR) of the {\em perturbed} eigenstate(s) (single one in case (i) or an 
average over all in case (iii)) in terms of the {\em unperturbed} eigenstates, and is thus directly related to the
{\em localization} properties of eigenstates of $U_\delta$ in terms of eigenstates of $U$.
However, except for the pathological case of initial state being a small combination of eigenstates of $U$ with weak perturbation, the
fidelity fluctuation is always between the limiting values ${\cal N}^{-1/2}$, and $\sqrt{3}/{\cal N}$. For an 
illustrations of transitions between weak and strong perturbations in case (iii) of a kicked top, see fig.~\ref{fig:Frms}.

Therefore, fidelity will decay according to one of the asymptotic formulae 
(\ref{eq:Fmix},\ref{eq:Fgauss},\ref{eq:Fsqrt},\ref{eq:Fcoh}) until it reaches the value of finite size fluctuation. 
This condition determines the time-scale $t_* = t_*({\cal N})$,
\begin{equation}
F_{{\cal N}\to\infty}(t_*({\cal N})) = F_{\rm rms}({\cal N}).
\label{eq:tstar}
\end{equation}
In non-ergodic and non-mixing regime this is the only relevant finite size time-scale, while in the mixing regime the 
situation is more complicated.

\subsubsection{Regime of ergodicity and fast mixing.}

Here, combining (\ref{eq:tstar}) with an exponential decay (\ref{eq:Fmix}) we get 
\begin{equation}
t_* = \mu \tau_{\rm em}\ln{\cal N} \approx \frac{\mu \hbar^2 d}{\delta^2\sigma_{\rm cl}}\ln(1/\hbar)
\end{equation}
where $\mu=-\ln F_{\rm rms}/\ln {\cal N}$ is a parameter which typically
lies in the range $1/2 < \mu < 1$ (except if the initial state is non-random
{\em and} the dynamics is non-ergodic (e.g. case (i)) 
{\em and} the perturbation is small $\delta \ll \delta_{\rm rm}$)
and depends on IPR and  the statistical operator $\rho$ as discussed above, 
and $\sigma_{\rm cl}$ is the classical limit of the transport coefficient (\ref{eq:sigma}).

The second new time-scale is related the to asymptotic non-decay of time correlations for finite ${\cal N}$ 
quantum dynamics, namely even if the system is classically mixing the quantum correlation function will have 
a small non-vanishing ($\hbar$-dependent) time average
\begin{equation}
\bar{C} = \overline{\ave{A(t)A(t')}} = \sum_n \rho_{nn} A_{nn}^2
\end{equation}
where $A_{mn} = \bra{\phi_m}A\ket{\phi_n}$. However, since the classical system is ergodic and mixing, 
we will use a version of {\em quantum chaos conjecture} 
saying that \cite{FeingoldPeres} then $A_{mn}$ are independent gaussian random variables with 
a variance given by the Fourier transformation $S(\omega)$ of the corresponding classical correlation 
function $C_{\rm cl}(t)$ at frequency $\omega=\phi_m-\phi_n$. On the diagonal we have $\omega=0$ and additional factor of $2$
due to random matrix measure on the diagonal (see e.g. \cite{Haake1}), so we can write
\begin{equation}
\bar{C} = \frac{4\sigma_{\rm cl}}{\cal N},
\label{eq:finitesize}
\end{equation}
where $\sigma_{\rm cl} = S(0)/2$ is the classical limit of (\ref{eq:sigma}).
Due to ergodicity, for large ${\cal N}$, this does not depend on the statistical operator $\rho$; neither in cases (ii,iii), nor
in case (i) if one assumes additional ensemble averaging.
The decay of fidelity (\ref{eq:FC}) will start to be dominated by the average plateau (\ref{eq:finitesize})
at time $t$, when
$\sum_{t'=0}^{t-1}{(t-t')\bar{C}} \approx 2\sigma_{\rm cl} t^2/{\cal N} \ge \sigma_{\rm cl} t$,
i.e. for times $t$ greater than $t_{\rm p}$
\begin{equation}
t_{\rm p}=\frac{1}{2}{\cal N} \propto \hbar^{-d}
\label{eq:tp}
\end{equation}
which is just the {\em Heisenberg time} associated to the inverse density of states.

\begin{figure}
\centerline{\includegraphics[width=2.5in]{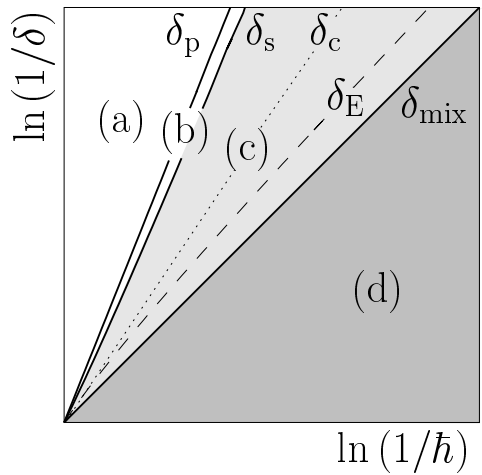}}
\caption{Schematic view of different regimes (a-d) of 
fidelity decay.}
\label{fig:delta}
\end{figure}

Now, depending on the interrelation among four (or five) time-scales 
$\tau_{\rm em} \propto \hbar^2 \delta^{-2}$,
$t_{\rm p} \propto \hbar^{-d} \delta^0$,
$t_* \propto \hbar^2 \ln(1/\hbar) \delta^{-2} d$,
$t_{\rm mix} \propto \hbar^0 \delta^0$,
(and $t_{\rm E} \propto \ln(1/\hbar)\delta ^0$ if we are considering coherent 
initial states, like e.g. \cite{Jalabert,Pastawski,Beenakker})
we can have four (or five) different regimes depending on three main scaling parameters: 
perturbation strength $\delta$, Planck's constant $\hbar$, and dimensionality $d$. 
Note that we always have $t_* > \tau_{\rm em}$. All different regimes can be reached by changing only 
$\delta$ while keeping $\hbar$ and $d$ fixed (see fig.~\ref{fig:delta}):
\\\\
({\bf a}) For sufficiently small perturbation $\delta$ we will 
have $t_{\rm p} < \tau_{\rm em}$. 
This means that $F_{\rm em}(t_{\rm p}) \approx 1$ and we will have initially quadratic decay (\ref{eq:Fr2}) with 
$\bar{C}$ given by an average finite size plateau (\ref{eq:finitesize}). 
This will occur for $\delta \le \delta_{\rm p}$ where
\begin{equation}
\delta_{\rm p}= \hbar \left(\frac{2}{\sigma_{\rm cl}{\cal N}}\right)^{1/2} =
\frac{\sqrt{2}(2\pi)^{d/2}}{({\cal V}\sigma_{\rm cl})^{1/2}}\hbar^{d/2+1}.
\label{eq:deltap}
\end{equation}
In fact, in this regime, also referred to \cite{Beenakker,Tomsovic} as {\em perturbative}, 
one may use a first order stationary perturbation theory on the eigenstates of $U_\delta$, yielding
$\phi^\delta_n = \phi_n + A_{nn}\delta/\hbar + {\cal O}(\delta^2)$, 
$V_{mn} = \delta_{mn} + {\cal O}(\delta)$, and rewrite (following \cite{Tomsovic}) the
finite size fidelity (\ref{eq:Ftfin}) in terms of a Fourier transform of a probability distribution $w(A)$ of 
diagonal matrix elements $A_{nn}$, $F_{\rm p}(t) = \int dA w(A) \exp(-iAt\delta/\hbar)$.
Since $w(A)$ is conjectured to be gaussian for classically ergodic and mixing system \cite{FeingoldPeres}, 
it follows that $F_{\rm p}(t)$ is also a gaussian with a semiclassically long time-scale $\tau_{\rm p}$
\begin{equation}
F_{\rm p}(t) = \exp{(-(t/\tau_{\rm p})^2/2)},\quad
\tau_{\rm p}=
\left(\frac{\cal N}{\sigma_{\rm cl}}\right)^{1/2}\frac{\hbar}{2\delta} =
\frac{{\cal V}^{1/2}}{(2\pi)^{d/2}\sigma_{\rm cl}^{1/2}}\frac{\hbar^{1-d/2}}{2\delta}.
\label{eq:Fp} 
\end{equation}
({\bf b}) If $\tau_{\rm em} < t_{\rm p} < t_*$ we will have a crossover from initial exponential decay of fidelity (\ref{eq:Fmix})
to a gaussian decay (\ref{eq:Fp}) at $t\sim t_{\rm p}$, which will terminate and go over to fluctuating behavior
when $F_{\rm p}(t) = F_{\rm rms}({\cal N})$. Note that this will happen before time 
$t_*$ which is estimated (\ref{eq:tstar}) based on a slower exponential decay (\ref{eq:Fmix}).
This regime will exist in perturbation range $\delta_{\rm p} < \delta < \delta_{\rm s}$ with an upper border $\delta_{\rm s}$ 
determined by the condition $t_{\rm p}=t_*$ 
\begin{equation}
\delta_{\rm s}=(\mu\ln{\cal N})^{1/2} \delta_{\rm p}.
\label{eq:deltas}
\end{equation} 
({\bf c}) If we still further increase $\delta$, we have the most 
interesting, 'fully nonperturbative' regime,
when $\tau_{\rm em} < t_* < t_{\rm p}$ and we will have a full exponential decay (\ref{eq:Fmix}), up to time $t_*$ 
when the fidelity reaches finite size fluctuations.
This regime continues for $\delta < \delta_{\rm mix}$ where the border
\begin{equation}
\delta_{\rm mix} = \frac{\hbar}{\sqrt{\sigma_{\rm cl}t_{\rm mix}}} = 
\left(\frac{\cal N}{2t_{\rm mix}}\right)^{1/2} \delta_{\rm p}.
\label{eq:deltam}
\end{equation}
is determined by the condition $\tau_{\rm em} = t_{\rm mix}$ which is a point
where the arguments leading to the factorization (\ref{eq:factor})
and exponential decay (\ref{eq:Fmix}) are no longer valid.
We note that the relative size of this window range
$\delta_{\rm mix}/\delta_{\rm s}=\sqrt{{\cal N}/(t_{\rm mix}\ln{\cal N})}$ increases both, 
in the semiclassical and in the thermodynamic limit. 
This regime also corresponds to `Fermi golden rule decay'
discussed in \cite{Beenakker}.
\\\\ 
({\bf d}) Further increasing $\delta > \delta_{\rm mix}$, the estimated 
fidelity decay time eventually becomes smaller than the classical mixing time 
$t_{\rm mix}$, $\tau_{\rm em} < t_{\rm mix}$. In this regime, the perturbation
is simply too strong so that the fidelity
effectively decays within the shortest observable time-scale ($t_{\rm mix}$).

However, if we consider a non-random, e.g. {\em coherent initial state} then
the quantum correlation function relaxes on a slightly longer, namely Eherenfest 
time-scale $t_{\rm E}$ so the regime (c) should terminate already at a little 
smaller upper border $\delta=\delta_{\rm E}$ which is naturally determined by 
the condition $\tau_{\rm em} = t_{\rm E}$
\begin{equation}
\delta_{\rm E} \approx \hbar\frac{\lambda^{1/2}}{[\sigma_{\rm cl}\ln(1/\hbar)]^{1/2}} \sim \frac{\delta_{\rm mix}}{[\ln(1/\hbar)]^{1/2}}.
\end{equation}
For coherent initial states one thus obtains an extra but very narrow regime 
$\delta_{\rm E} < \delta < \delta_{\rm mix}$ (describing the
time-range $t_{\rm mix} < t < t_{\rm E}$) where the fidelity decay can be
computed in terms of classical Lyapunov exponents \cite{Jalabert,Pastawski}.

\subsubsection{Non-mixing and non-ergodic regime.}

In this regime things are simpler, as we do not have to worry about the average plateau in the correlation function due to a 
finite ${\cal N}$ because we already have a higher average time correlation $\bar{C}\to \bar{C}_{\rm cl}\neq 0$ (\ref{eq:Cinfty}) due to 
non-mixing nature of dynamics. Thus we have here only two relevant time-scales, namely $\tau_{\rm ne}$ giving initial quadratic decay 
(\ref{eq:Fr2}), and the finite size fluctuation time-scale $t^*$ (\ref{eq:tstar}) which depends on the properties of the
initial state (power law (\ref{eq:Fsqrt}) for a random initial state, versus gaussian (\ref{eq:Fcoh}) for a coherent initial state).
We conjecture, based on a rigorous result in spin $1/2$ chains \cite{Prosen01}, that the
fidelity decay in the thermodynamic limit $d\to\infty$ will generically approach a gaussian (\ref{eq:Fgauss}), which 
is consistent with increasing power $d/2$ of power-law fidelity decay (\ref{eq:Fsqrt}) for finite dimension $d$.
\\\\
We can summarize our findings by the following question:
Can we have a regime with $\tau_{\rm ne} < \tau_{\rm em}$, so that the fidelity will decay faster for a 
``regular'' (non-ergodic) than for a ``chaotic'' (ergodic and mixing) dynamics? Indeed, if 
\begin{equation}
\!\!\!\!\!\!\!\!\!\!\!\!\delta <
\left\{
\begin{array}{ll}
\delta_{\rm r}:=\hbar\bar{C}_{\rm cl}^{1/2}/\sigma_{\rm cl} \;\propto\; {\rm \delta_{\rm mix}} \;\propto\; \hbar      
& \hbox{random initial state} \\ 
\delta_{\rm c}:=\hbar^{3/2}(\ve{a}'\cdot\Lambda^{-1}\ve{a}')^{1/2}/(\sqrt{2}\sigma_{\rm cl}) \;\propto\; \hbar^{3/2}
& \hbox{coherent initial state}
\end{array}
\right. 
\label{eq:delta_con}
\end{equation}
where $\sigma_{\rm cl}$ and $\bar{C}_{\rm cl}$ are computed for a 
mixing and non-ergodic classical dynamics, respectively, 
then the decay in a non-ergodic case will be faster than in a mixing case. 
We can see that the condition (\ref{eq:delta_con}) can be generally satisfied for a random initial state, while for
a coherent initial state it can be satisfied above the (second) perturbative border (see fig.~\ref{fig:delta}), 
$\delta_{\rm c} > \delta > \delta_{\rm s}$, only in more than one dimension $d > 1$.
We note that our result is not contradicting any of the existing findings on quantum-classical correspondence.
For example, a growth of quantum dynamical entropies \cite{Alicki} persists only up to logarithmically short Ehrenfest time 
$t_{\rm E}$, which is the upper bound for the validity of another semiclassical approach to the fidelity decay 
\cite{Jalabert,Pastawski} and within which one would always find $F^{\rm coh}_{\rm ne}(t) > F_{\rm em}(t)$ 
above the perturbative border $\delta > \delta_{\rm p}$, whereas our theory reveals new nontrivial 
quantum phenomena with a semiclassical prediction (but not correspondence!) much beyond that
time. If we let $\hbar\to 0$ first, and then $\delta \to 0$, i.e. we keep $\delta \gg \delta_{\rm r,\rm c}(\hbar)$,
then we recover a result supported by a classical intuition, namely that the regular (non-ergodic) dynamics
is more stable than the chaotic (ergodic and mixing) dynamics. On the other hand, if
we let $\delta\to 0$ first, and only after that $\hbar\to 0$, i.e. satisfying (\ref{eq:delta_con}),
we find somewhat counterintuitive results saying that chaotic (mixing) dynamics is more stable than the regular one.
We can conclude the section by saying that we have {\em three non-commuting limits}, namely 
the {\em semiclassical limit} $\hbar \to 0$ 
the {\em perturbation strength} limit $\delta \to 0$, 
and the {\em thermodynamic limit} $d \to \infty$, 
such that no pair of these limits commutes.

\section{Numerical experiments}

\subsection{Kicked top}

Here we wish to verify and demonstrate the results of the previous section by numerical experiments. 
For this purpose we choose the Haake's quantized kicked top \cite{Haake2} since this model served as a 
model example for many related studies \cite{Shack94,Beenakker,Alicki,Fox,Casati,Breslin,Haake00}.
The unitary propagator reads
\begin{equation}
U = U(\alpha,\gamma)=\exp{(- i \gamma J_{\rm y})} \exp{(-i \alpha J_{\rm z}^2/2J)},
\label{eq:Ukt}
\end{equation}
where $J_k$ ($k={\rm x,y,z}$) are quantum angular momentum operators, 
$[J_k,J_l]=i \epsilon_{klr} J_r$. (Half)integer $J$ determines the size of the Hilbert space $2J+1$ and
the value of effective Planck's constant $\hbar=1/J$.
The perturbation is defined by perturbing the parameter $\alpha$, $U_\delta=U(\alpha+\delta,\gamma)$, 
so that the generator $A$ is
\begin{equation}
A=\frac{1}{2}\left(\frac{J_{\rm z}}{J}\right)^2,
\label{eq:KTa}
\end{equation}
Physically, the system (\ref{eq:Ukt}) represents a twist around ${\rm z}$-axis 
followed by a rotation for an angle $\gamma$ around ${\rm y}$-axis. The classical limit is obtained by letting 
$J = 1/\hbar \to \infty$ and writing the classical angular momentum in terms of a unit vector on a sphere 
$\ve{r} = (x,y,z) = \ve{J}/J$.
The Heisenberg equation for the SU(2) operators $\ve{J}$, $\ve{J}' = U^\dagger \ve{J} U$, reduces to the classical
area preserving map of a sphere
\begin{eqnarray}
x'&=& \cos{\gamma}(x \cos{\alpha z}-y \sin{\alpha z})+z\sin{\gamma} \nonumber \\
y'&=& y \cos{\alpha z} + x \sin{\alpha z} \nonumber \\
z'&=& z \cos{\gamma}+\sin{\gamma}(y\sin{\alpha z}-x\cos{\alpha z}).
\label{eq:KTclass}
\end{eqnarray}
Note that in the classical limit the perturbation generator is
\begin{equation}
a(\ve{r})=\frac{z^2}{2}.
\label{eq:Aclass}
\end{equation}
For $\alpha=0$ the system is integrable, while with increasing $\alpha$ there is a transition to chaotic motion. 
The second parameter $\gamma$ is usually set to $\pi/2$, however in our numerical simulation we will use two different 
values exhibiting qualitatively different correlation decay (for large $\alpha$): the 'standard' case $\gamma=\pi/2$ where 
$C_{\rm cl}(t)$ decays in oscillatory way and the case $\gamma=\pi/6$ where $C_{\rm cl}(t)$ decays monotonically
(see fig.~\ref{fig:class30}).
\par
In the case of $\gamma=\pi/2$ we have two discrete symmetries. The evolution $U$ 
commutes with $R_{\rm x}$ and $R_{\rm y}$, the rotations for $\pi$ around ${\rm x}$ and ${\rm y}$ axes, respectively. 
The Hilbert space is therefore reducible into three invariant subspaces (using notation of Peres's book \cite{Peres1} with the
basis $\ket{m}$ of eigenstates of $J_{\rm z}$ and assuming $J$ to be an {\em even integer}): 
EE of dimension $J/2+1$ with the basis states $\ket{0}$ and $\{ \ket{2m}+\ket{-2m} \}/\sqrt{2}$; 
OO of dimension $J/2$ with the basis $\{ \ket{2m-1}-\ket{-(2m-1)} \}/\sqrt{2}$; 
OE of dimension $J$ with the basis $\{ \ket{2m}-\ket{-2m}\}/\sqrt{2}$ and $\{ \ket{2m-1}+\ket{-(2m-1)}\}/\sqrt{2}$ 
with $m$ in all three cases running through $m=1,\ldots,J/2$. For $\gamma \neq \pi/2$ the spaces OO and EE coalesce as 
$R_{\rm y}$ is the only discrete symmetry left.
In numerical experiments we always choose OE subspace so that the dimension of the Hilbert space is ${\cal N}=J$. 
\par
We will compute the fidelity using three different statistical operators $\rho$: (1) $\rho=\mathbbm{1}/{\cal N}$ corresponding
to full Hilbert space average, (2) pure random initial state $\rho=\ket{\psi}\bra{\psi}$ 
(components $c_m = \braket{m}{\psi}$ being independent gaussian pseudo-random numbers) giving the same results as (1) however with a
higher finite size fluctuation plateau $F_{\rm rmt}$ (as discussed in subsect.~\ref{sec:timepert}), 
and (3) pure minimal wavepacket initial state 
$\rho=\ket{\vartheta,\varphi}\bra{\vartheta,\varphi}$, namely SU(2) coherent state $\ket{\vartheta,\varphi}$
centered at the point $\ve{n}=(\sin{\vartheta}\cos{\varphi},\sin{\vartheta}\sin{\varphi},\cos{\vartheta})$ on a unit sphere
\begin{equation}
\ket{\vartheta,\varphi}=\sum_{m=-J}^{J}{\left( {2J \atop J+m} \right)^{1/2} 
\cos{(\vartheta/2)}^{J+m} \sin{(\vartheta/2)}^{J-m} e^{-i m \varphi} \ket{m}}.
\label{eq:SU2coh}
\end{equation}

\subsubsection{Mixing regime.}

In this regime we will demonstrate two main different decays of fidelity as discussed in subsect.~\ref{sec:timepert}, 
namely the exponential decay $F_{\rm em}(t)$ (\ref{eq:Fmix}) for $\delta_{\rm s} < \delta < \delta_{\rm mix}$ 
and a finite size (perturbative) gaussian decay $F_{\rm p}(t)$ (\ref{eq:Fp})
for a small perturbation $\delta < \delta_{\rm p}$. 
We choose large $\alpha=30$ to ensure fast mixing. 
This value seems to be excessively large, but for smaller $\alpha$ we still have ``sticky'' structures in 
classical phase space which in turn cause slow algebraic tails in the classical correlation function. 
We must remember that the derivation of an exponential decay required well defined mixing time-scale 
$t_{\rm mix}$ which is not the case if we have a slow power-law decay of correlations. 
\par
Exponential decay time $\tau_{\rm em}$ (\ref{eq:Fmix}) is determined by an integral/sum of 
time correlation function (\ref{eq:sigma}), which can be calculated in the semiclassical regime ($\sigma_{\rm cl}$)
by means of the correlation function of the classical map
\begin{equation}
C_{\rm cl}(t)=\frac{1}{4}\ave{\tilde{z}^2(0) \tilde{z}^2(t)}.
\label{eq:Cclass}
\end{equation}
$\tilde{z}^2=z^2-\ave{z^2}$ is a ``traceless'' perturbation where averaging over the sphere gives $\ave{z^2}=1/3$. 
\begin{figure}
\centerline{\includegraphics{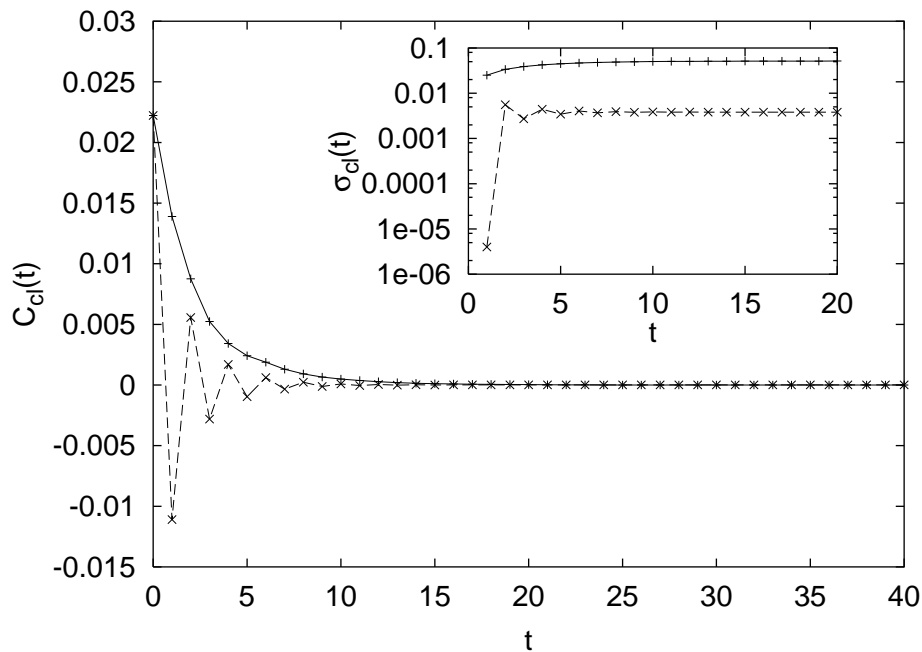}}
\caption{Classical correlation function $C_{\rm cl}(t)$ (\ref{eq:Cclass}) for $\alpha=30$, and $\gamma=\pi/6$ 
(top, full curve) and $\gamma=\pi/2$ (bottom, broken curve). Finite time integrated correlation function is shown in the inset,
converging to $\sigma_{\rm cl}=0.00385$ and $0.0515$, for $\gamma=\pi/2$ and $\pi/6$, respectively. 
Averaging over $10^5$ random initial conditions on a sphere is performed.}
\label{fig:class30}
\end{figure}
The classical correlation functions calculated by using the classical map (\ref{eq:KTclass}) are shown in fig.~\ref{fig:class30}. 
For $\gamma=\pi/2$ the correlation function is oscillating with an exponential envelope hence the transport coefficient 
$\sigma_{\rm cl}=0.00385$ is quite small. 
For $\gamma=\pi/6$ the correlation decay is monotonic and exponential with $\sigma_{\rm cl}=0.0515$. 
The decay of quantum fidelity (\ref{eq:Fmix}) can now be obtained by using the {\em classical} limit 
$\sigma\to\sigma_{\rm cl}$:
\begin{equation}
F_{\rm em}(t)=\exp{(-\delta^2 J^2 \sigma_{\rm cl} t)},
\label{eq:Fmclass}
\end{equation}
This formula has been compared with the exact numerical calculation of fidelity (\ref{eq:Ft}) where 
averaging over the whole Hilbert space has been employed, i.e. $\rho=\mathbbm{1}/{\cal N}$, and, 
as expected due to ergodicity, there was {\em no difference} for sufficiently large $J$ when we have chosen 
a fixed coherent initial state. As the finite size fidelity fluctuation level $F_{\rm rms-iii}$ 
(\ref{eq:Frmsc}) decreases with increasing Hilbert space dimension, 
we chose large $J=4000$ in order to be able to check exponential decay (\ref{eq:Fmclass}) 
over as many orders of magnitude as possible. 
\begin{figure}
\centerline{\includegraphics{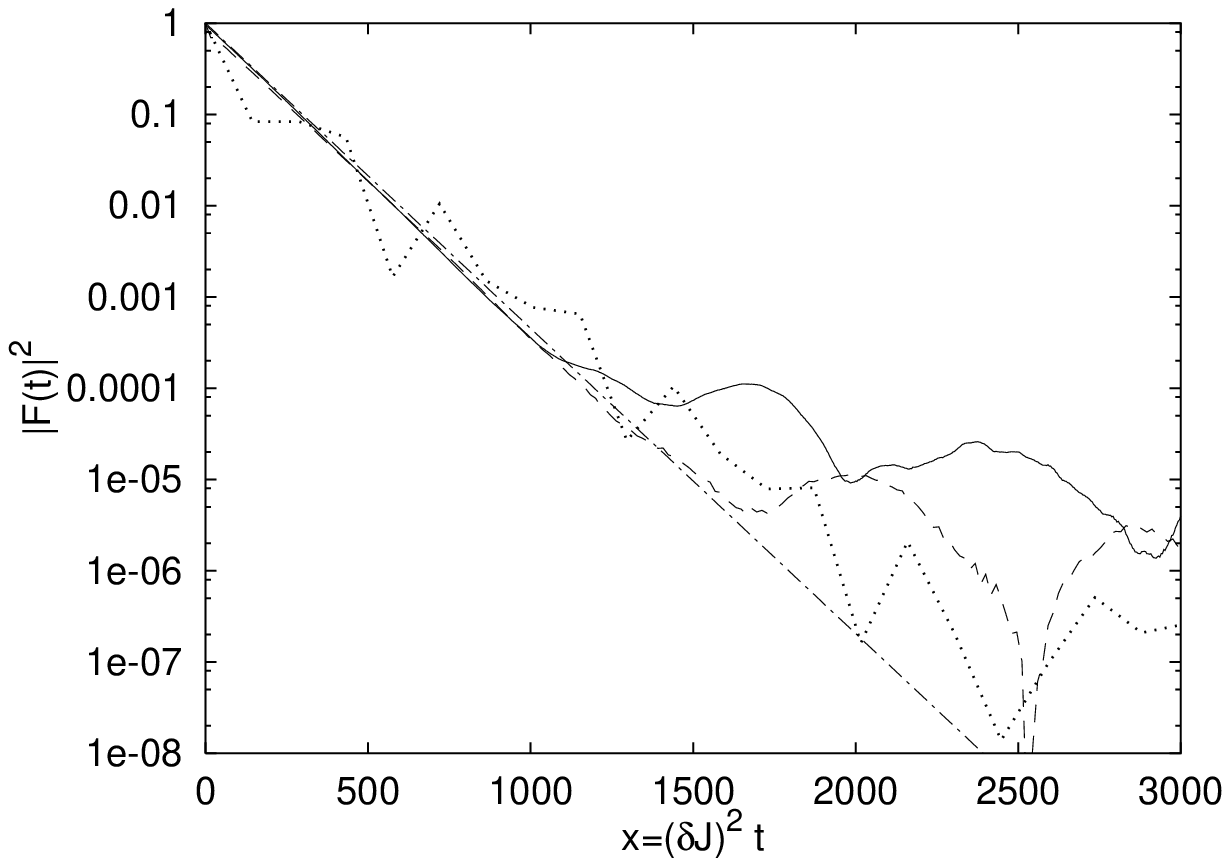}}
\centerline{\includegraphics{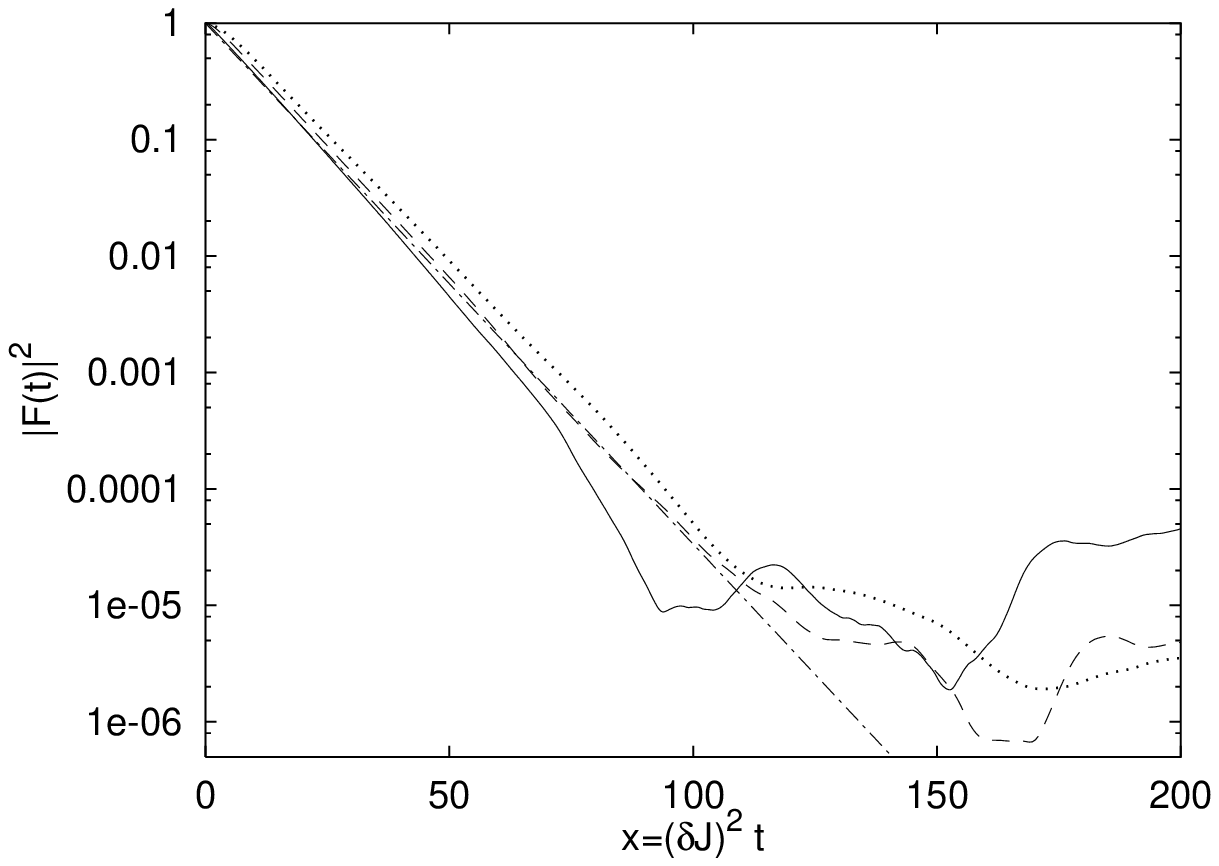}}
\caption{Quantum fidelity for a kicked top with parameters $\alpha=30$, $J=4000$ and full trace average. 
Top figure is for $\gamma=\pi/2$ and for $\delta=5\cdot 10^{-4},1\cdot 10^{-3},3\cdot 10^{-3}$ 
(solid, dashed, dotted curves, respectively)
Bottom figure is for $\gamma=\pi/6$ and $\delta=1\cdot 10^{-4},2\cdot 10^{-4},3 \cdot 10^{-4}$, 
(solid, dashed, dotted curves, respectively).
Chain line in both cases gives the theory (\ref{eq:Fmclass}) with classically computed $\sigma_{\rm cl}$.
Note that the largest $\delta=3\cdot 10^{-3}$ case in the top figure (dotted curve) corresponds to $\tau_{\rm}\approx 2$,
so it is already over the upper border of the regime ({\bf c}, subsect.~\ref{sec:timepert}) $\delta > \delta_{\rm mix}$ 
but the agreement with the theory (\ref{eq:Fmclass}) is still quite good, appart from oscillations. 
This is due to the oscillatory nature of time-correlations making the factorization assumption (\ref{eq:factor})
justified (on average) even for much smaller time $t$ as required.
}
\label{fig:j40k}
\end{figure}
The results are shown in fig.~\ref{fig:j40k}. The smallest and the largest $\delta$ shown roughly correspond to borders 
$\delta_{\rm s}$ and $\delta_{\rm mix}$, respectively. As we can see, the agreement with an exponential decay is excellent, 
at least over four decades in probability $|F(t)|^2$.
Note that for $\gamma=\pi/2$ and the largest $\delta=3\cdot 10^{-3}$ the time-scale of the decay of fidelity is comparable to the 
classical time-scale $t_{\rm mix}$ so the factorization assumption (\ref{eq:factor}) is strictly not applicable any more.
However, due to oscillatory correlation decay, overall agreement with the theory (\ref{eq:Fmclass}) is still rather good but the
oscillations in the correlation decay reflect in the oscillations of the fidelity decay (around the theoretical exponential curve).
Of course, one does not need such a large $J$ in order to have an exponential decay, but for smaller $J$ the fluctuation
level $F_{\rm rms}$ will be higher so the exponential decay (\ref{eq:Fmclass}) will persist for correspondingly smaller
time, namely up to $t_*(J)$ (\ref{eq:tstar}).  
\par
\begin{figure}
\centerline{\includegraphics{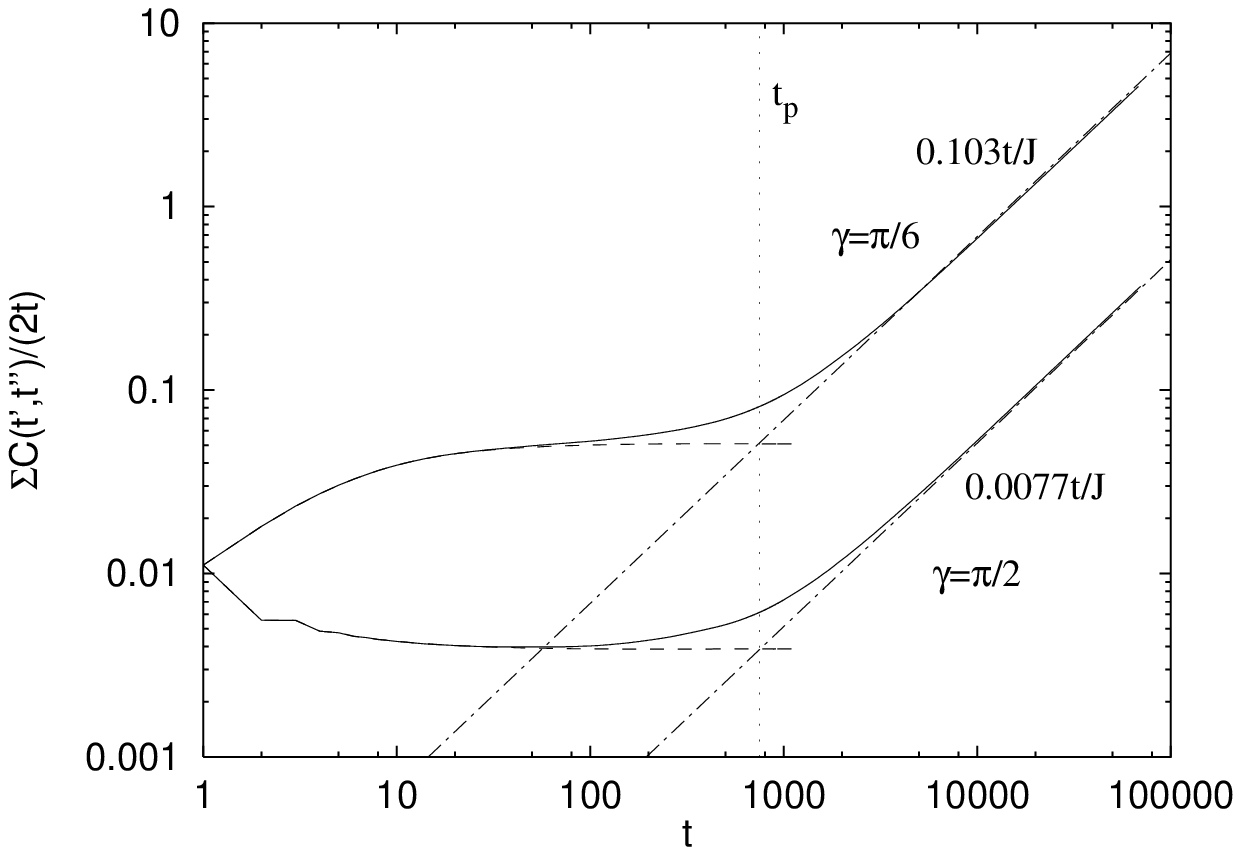}}
\caption{Finite time quantum correlation sum 
$\sigma(t)=\sum_{t',t''=0}^{t-1} {C(t',t'')/2t}$ (solid curves) 
and the corresponding classical sum 
$\sigma_{\rm cl}(t) = \sum_{t',t''=0}^{t-1}{C_{\rm cl}(t',t'')}/2t$ 
(dashed curves saturating at $\sigma_{\rm cl}$ and ending at $t \sim 1000$) for $\alpha=30,J=1500$. 
Upper curves are for $\gamma=\pi/6$ while lower curves are for $\gamma=\pi/2$. 
Chain lines are best fitting asymptotic linear functions corresponding to $\bar{C}t$, 
$0.0077t/J$ for $\gamma=\pi/2$ and $0.103 t/J$ for $\gamma=\pi/6$.}
\label{fig:cinf}
\end{figure}
Then we focus on the so-called perturbative regime $\delta < \delta_{\rm p}$ where the fidelity decay will be dictated by a 
finite size correlation average (\ref{eq:finitesize}), so according to eq. (\ref{eq:Fp})
\begin{equation}
F_{\rm p}(t)=\exp{(-2 \delta^2 J \sigma_{\rm cl} t^2)}.
\label{eq:Fpert}
\end{equation}  
We numerically computed $\bar{C}$ (\ref{eq:Cinfty}) for $J=1500$, $\alpha=30$ in order to show that it is given by 
the theoretical value (\ref{eq:finitesize}). Quantum correlation function has been computed 
$C(t',t'') = \ave{\tilde{A}_{t'}\tilde{A}_{t''}}$ by means of a traceless perturbation
$\tilde{A} = \frac{1}{2}(J_z/J)^2 - \frac{1}{12}[(2J+1)(J+1)/J^2]\mathbbm{1}$.
In fig.~\ref{fig:cinf} we show a finite time correlation sum $\sigma(t)=\frac{1}{2t}\sum_{t',t''=0}^{t-1} C(t',t'')$ which
exhibits a crossover, at the Heisenberg time $t_{\rm p}=J/2$, from the plateau given by $\sigma_{\rm cl}$ to a 
linear increase $\bar{C}t$ due to finite size correlation average (\ref{eq:finitesize}) $\bar{C}=4 \sigma_{\rm cl}/J$.
\begin{figure}
\centerline{\includegraphics{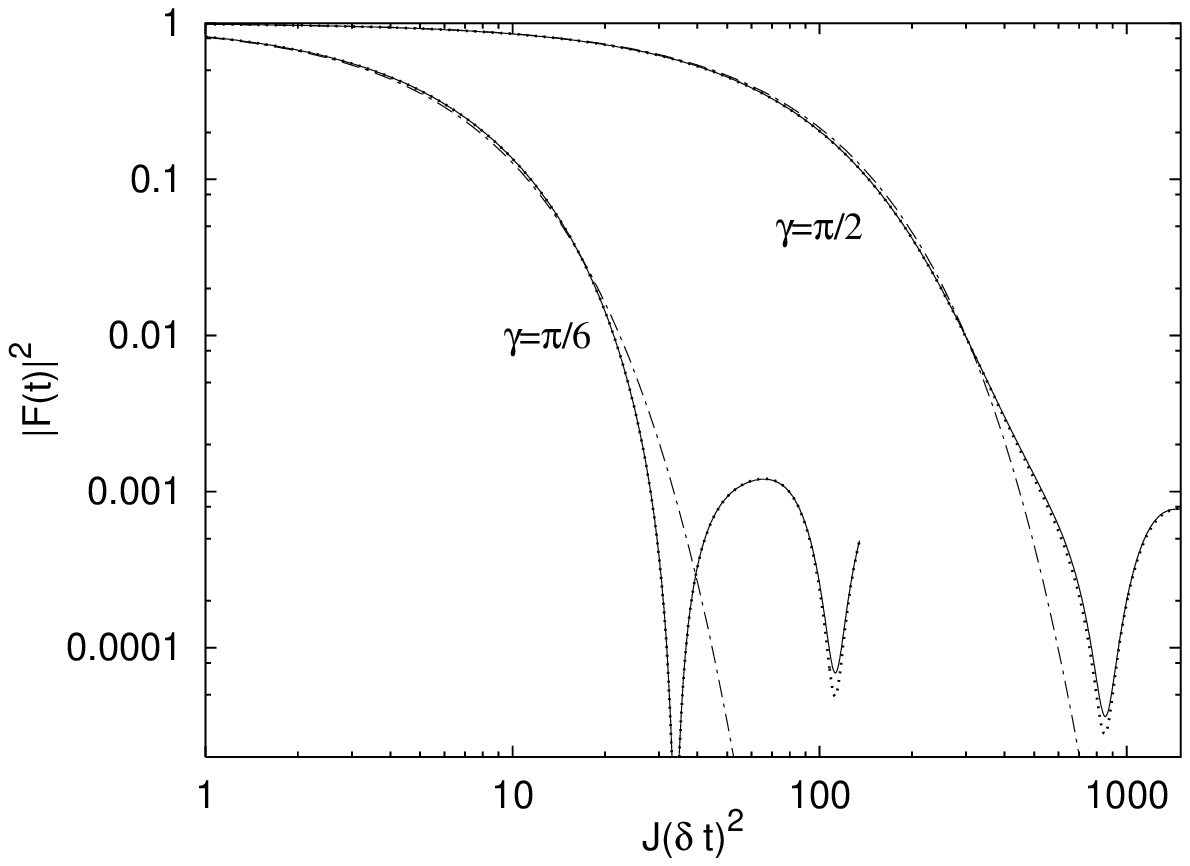}}
\caption{Quantum fidelity in the perturbative regime $\delta<\delta_{\rm p}$ for $\alpha=30$, $J=1500$, and 
$\gamma=\pi/2$ and $\pi/6$, calculated as a full trace Hilbert space average as a function of the scaled
variable $J(\delta t)^2$. For $\gamma=\pi/2$ data for $\delta=1\cdot 10^{-6}$ (solid curve) and $5\cdot 10^{-6}$ 
(dotted curve) are shown. For $\gamma=\pi/6$, $\delta=3\cdot 10^{-7}$ (solid) and $1\cdot 10^{-6}$ (dotted) are shown. 
Note that for both $\gamma$ the curves for both $\delta$ practically overlap. The chain curves are theoretical 
predictions (\ref{eq:Fpert}) with classically computed $\sigma_{\rm cl}$.}
\label{fig:j15kpert}
\end{figure}
The excellent agreement between prediction (\ref{eq:Fpert}) and full numerical calculation of fidelity is shown in 
fig.~\ref{fig:j15kpert}. 
In view of our findings this so-called \cite{Beenakker} perturbative regime can be understood as a simple consequence of a 
finite Hilbert space dimension. For times larger than the Heisenberg time 
$t_{\rm p}$ every quantum system behaves effectively as an integrable one, e.g. with a finite time average correlation
plateau.

\subsubsection{Non-mixing regime.}
In a kicked top this regime is realized for sufficiently small value of $\alpha$. 
If the classical phase space has a mixed (KAM) 
structure, non-mixing regime of fidelity decay may be obtained by choosing a localized initial state (e.g. coherent state) 
located in a regular part of the phase space. 
However, such a situation may easily lead to the opposite conclusion (as compared to generic situation) for insufficiently 
large dimension ${\cal N}$. As discussed in subsect.\ref{sec:timepert}, the fidelity fluctuation plateau is determined by 
the number of constituent propagator eigenstates $\ket{\phi_n}$ which are effectively needed to expand the initial state. 
For a coherent state sitting inside a (not too large) regular (KAM) island this number can be fairly small for numerically 
realizable Hilbert space dimensions, thus prohibiting any significant fidelity decay as observed in \cite{Peres2}.
Nevertheless, we would still see the initial quadratic decay in the linear response regime but we would not be able
to verify higher orders in the long-time expansion of fidelity.
Therefore, in order to make a situation numerically as clean as possible, we choose a small value of parameter $\alpha=0.1$, 
such that the classical dynamics is almost integrable and that the majority of phase space corresponds to regular motion so that
the number of constituent eigenstates for coherent states is as large as possible (on average).
\par
Here we focus only on the case $\gamma=\pi/2$. 
For small $\alpha$, the quantum and classical evolution is a (slightly perturbed) rotation around ${\rm y}$ 
axis and the time averaged perturbation can be computed analytically (in the leading order in $\alpha$)
\begin{equation}
\bar{A}=\frac{1}{4J^2}(J_{\rm z}^2+J_{\rm x}^2)=\frac{1}{4}\left(1-(J_{\rm y}/J)^2\right),\qquad
\bar{a}=\frac{1}{4}(1-y^2).
\label{eq:ave} 
\end{equation}
We will now use (approximate) analytical results for $\alpha\to 0$ to compare with
numerics for $\alpha=0.1$. It should be noted that our leading order analytical approximations could easily be systematically
improved using a classical perturbation theory (treating $\alpha$ as a perturbing parameter).
However, since the agreement, as shown below, is almost perfect in all cases, we see no need for refinement at this 
level.
\par
First, we consider the full trace average, $\rho=\mathbbm{1}/J$, and 
starting from the expression (\ref{eq:Favg}), $F(t)=\ave{\exp{(i t \bar{A}\delta/\hbar)}}$,
write the fidelity as a sum over all 
eigenvalues of $J_{\rm y}^2$, namely $(2m-1)^2$, for $m=1,\ldots,J/2$ (in OE subspace),
\begin{equation}
|F(t)|=\left|\frac{2}{J} \sum_{m=1}^{J/2}{\exp{(i \delta t (2m-1)^2/4J)}}\right|.
\label{eq:Fexact}
\end{equation}
For large $J$ we can replace the sum with an integral and get
\begin{equation}
|F(t)|=\sqrt{\frac{\pi}{\delta J t}} \left|{\rm erfi}(\frac{1}{2} e^{i \pi/4} \sqrt{\delta J t})\right|,
\label{eq:Ferfi}
\end{equation}
\begin{figure}
\centerline{\includegraphics{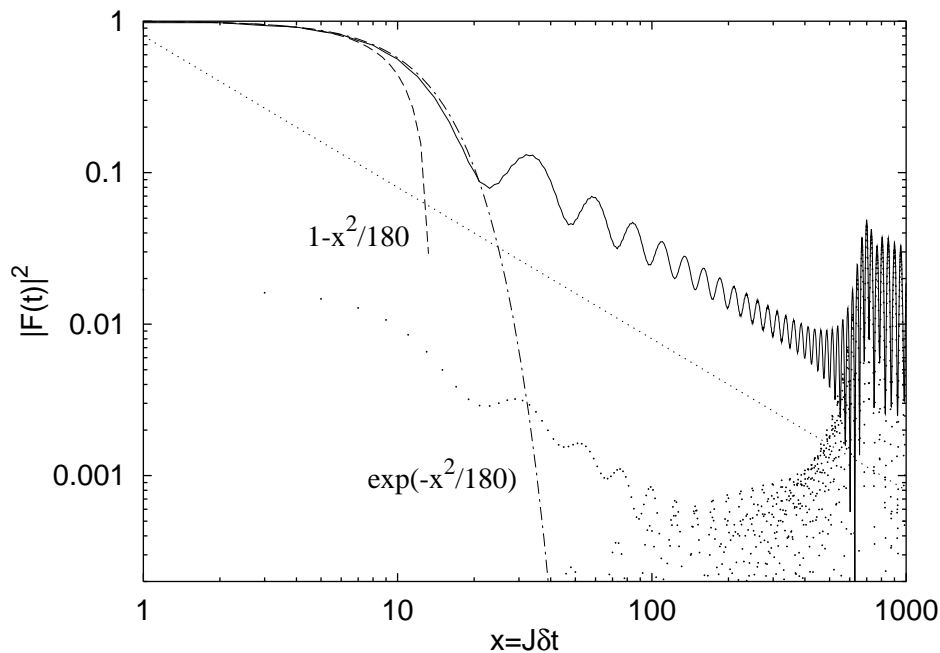}}
\caption{Fidelity in quasi-integrable regime for $\alpha=0.1$, $\gamma=\pi/2$, $\delta=0.01$, $J=100$, and 
$\rho=\mathbbm{1}/J$, in OE subspace. 
Solid curve gives the result of a numerical simulation. Isolated dots denote differences between numerical calculation and the 
analytic formula (\ref{eq:Ferfi}) for $\alpha\to 0$ 
$||F_{\rm num.}(t)|^2-|F_{\rm anali.}(t)|^2|$. 
The dotted line gives a predicted asymptotic decay $\propto t^{-1/2}$, and the dashed/chain curves are the 
predicted fidelity decays at small times, namely the second order expansion $|F(t)|^2=1-(Jt\delta)^2/180$, and 
'improved' by the gaussian (\ref{eq:Fgauss}).}
\label{fig:analit}
\end{figure}
where ${\rm erfi}(z)=\frac{2}{i\sqrt{\pi}}\int_0^{iz}{e^{-t^2} dt}$ is a complex error function with a limit 
$\lim_{x \to \infty}{|{\rm erfi}(\frac{1}{2} e^{i \pi/4} \sqrt{x})|}=1$ to which it approaches by oscillating 
around $1$. We therefore have an analytic expression for the fidelity (\ref{eq:Ferfi}) in the case of an uniform 
average over whole Hilbert space or, equivalently, over one random initial state. Its asymptotic decay is $t^{-1/2}$ which 
agrees with the general semiclassical asymptotics (\ref{eq:Fsqrt}) and we expect initial quadratic decay 
(\ref{eq:Fr2}) for small times $t < \tau_{\rm ne}$
Decay rate $\tau_{\rm ne}$ is determined by the time averaged correlation $\bar{C}$ (\ref{eq:Cinfty}) which can be
calculated explicitly in the limit $\alpha\to 0$ where the classical correlation function $C_{\rm cl}(t)$ alternates 
for even/odd times as
$C_{\rm cl}(2t)=\ave{\tilde{z}^2(0) \tilde{z}^2(2t)}/4=-1/90$, 
$C_{\rm cl}(2t+1)=\ave{\tilde{z}^2(0) \tilde{z}^2(2t+1)}/4=1/45$,
giving
\begin{equation}
\bar{C}_{\rm cl}\vert_{\alpha=0} = \frac{1}{2}\left(-\frac{1}{90}+\frac{1}{45}\right)=\frac{1}{180},
\label{eq:Cinftykt}
\end{equation}
and the fidelity is expected to decay as (\ref{eq:Fr2}) with 
$\tau_{\rm ne} = \sqrt{180}/(J\delta)$,
for short times, $t < \tau_{\rm ne}$. The short-time formula (\ref{eq:Fr2}) and the full analytic expression (\ref{eq:Ferfi}) 
are compared with the numerical simulation in fig.~\ref{fig:analit}. The agreement is very good and, surprisingly enough, 
the gaussian approximation (\ref{eq:Fgauss}) for small times 
is observed to be valid considerably beyond the second order expansion (\ref{eq:Fr2}). 
Quite interesting is the regime where the decay time $\tau_{\rm ne}=\sqrt{180}/(J\delta)$ for a ``regular'' 
dynamics will be smaller than a decay time $\tau_{\rm em}=1/(\delta^2 J^2 \sigma_{\rm cl})$ (\ref{eq:Fmclass}) for a ``chaotic'' 
dynamics. This will happen for $\delta < \delta_{\rm r} =  1/(J \sigma_{\rm cl} \sqrt{180})$ (\ref{eq:delta_con}). 
This border has the  same scaling with $J$ as $\delta_{\rm mix}$ (\ref{eq:deltam}).    
\par
\begin{figure}
\centerline{\includegraphics{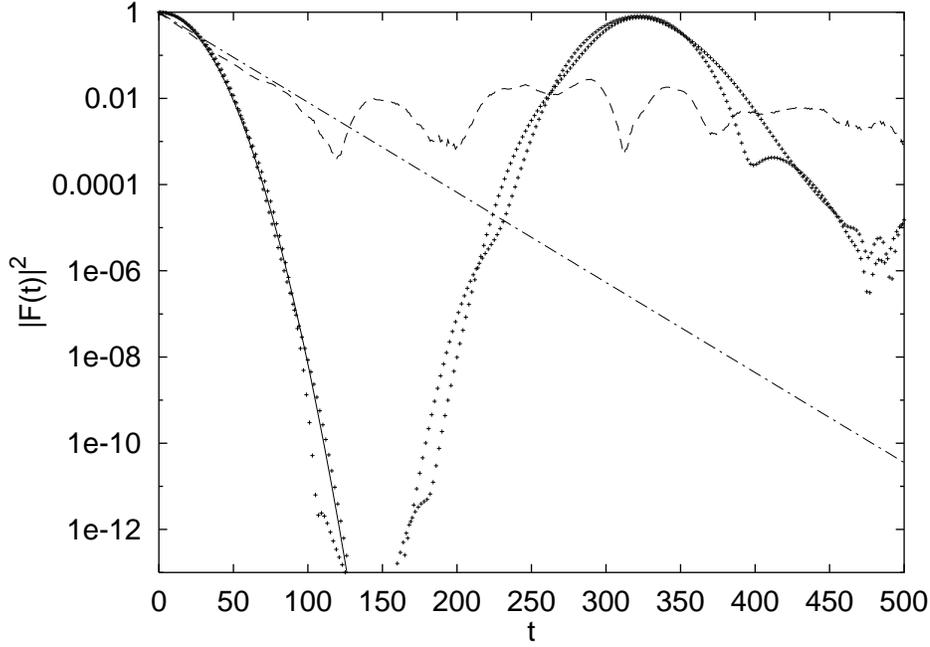}}
\caption{Fidelity for $\gamma=\pi/2$, $\delta=0.025$ and $J=100$ on OE subspace.
The dashed curve is a simulation for $\alpha=30$ (mixing regime, full trace average). 
The pluses are for a pure coherent initial state (see text for details) at $\alpha=0.1$ (non-mixing regime). 
The chain and solid curves are, respectively, the theoretical exponential (\ref{eq:Fmclass}) and gaussian (\ref{eq:Fktcoh}) decays.}
\label{fig:parad1}
\end{figure}

Second, considering SU(2) coherent initial state (\ref{eq:SU2coh}), 
we could proceed along the same line, namely by an analytic calculation. But rather than that, we will illustrate the usefulness 
of a semiclassical formula for $F_{\rm ne}^{\rm coh}(t)$ (\ref{eq:Fcoh}). This is a more general approach, as the explicit 
analytical calculation is usually not possible. Let us by $\tilde{\vartheta},\tilde{\varphi}$ denote the spherical angular
coordinates measured with respect to the y-axis. Then $(I=\cos\tilde{\vartheta}=y,\tilde{\varphi})$ represent canonical 
action-angle coordinates for the integrable case $\alpha\to 0$. Furthermore, the coherent state (\ref{eq:SU2coh}) acquires 
a semiclassical gaussian form (\ref{eq:CS}) in the EBK basis $\ket{n}$,
$J_{\rm y}\ket{n} = n\ket{n}$, namely
\begin{equation}
|\braket{n}{\tilde{\vartheta},\tilde{\varphi}}|^2 
\propto \exp{\left(-\frac{(n\hbar-\cos{\tilde{\vartheta}})^2}{\hbar\sin^2{\tilde{\vartheta}}} \right)},\quad \hbar=\frac{1}{J}.
\end{equation}
The squeezing parameter $\Lambda$ reads 
\begin{equation}
\Lambda=1/\sin^2{\tilde{\vartheta}}=1/(1-y^2).
\end{equation}
In order to apply the general formula (\ref{eq:Fcoh}) we need to express the classical time average (\ref{eq:ave}) 
in terms of a canonical action, $\bar{a}(I) = (1-I^2)/4$, and evaluate the derivative,
$|\partial \bar{a}(I)/\partial I|^2=\frac{1}{4} I^2$, giving
$|F^{\rm coh}_{\rm ne}(t)|=\exp{(-\delta^2 J t^2 I^2 (1-I^2)/16)}$. Rewriting this expression in terms of original spherical
angles $\vartheta$ and $\varphi$ measured with respect to ${\rm z}$-axis, we obtain
\begin{equation}
F^{\rm coh}_{\rm ne}(t)=\exp{\left(-\frac{\delta^2 J t^2}{64} 
\{\sin^2{2\vartheta}\sin^2{\varphi}+\sin^4{\vartheta} \sin^2{2\varphi} \}
\right)}.
\label{eq:Fktcoh}
\end{equation}
Here the same interesting question, namely when do we have $\tau_{\rm ne-coh} < \tau_{\rm em}$, 
results in a condition (\ref{eq:delta_con}) $\delta < 1/(4\sigma_{\rm cl} \sqrt{2J^3})=
\delta_{\rm s}/\sqrt{32 \sigma_{\rm cl} \ln{J}}$ (we did not write the factor involving trigonometric functions (\ref{eq:Fktcoh}) 
which is of order $1$). 
This condition generally can not be meet if we also require $\delta > \delta_{\rm s}$ (in order to see full exponential decay in the mixing regime), except if $32 \sigma_{\rm cl} \ln{J} < 1$. This is for instance satisfied for $J=100$ and $\gamma=\pi/2$, for 
which $\sigma_{\rm cl}=0.00385$. We checked this by a numerical simulation and the result for a coherent state centered at 
$(\vartheta,\varphi)=\pi(1/\sqrt{3},1/\sqrt{2})$, 
for which $\sin^2{2\vartheta}\sin^2{\varphi}+\sin^4{\vartheta} \sin^2{2\varphi}=0.96$, is shown in fig.~\ref{fig:parad1}.
We can see that for $t > 50$ the fidelity in a non-mixing regime ($\tau_{\rm ne-coh}=23$) is lower than the fidelity in a mixing regime ($\tau_{\rm em}=42$). For larger times, $t > t^*$, non-mixing decay $F^{\rm coh}_{\rm ne}$ 
displays revivals of fidelity. 
\par
\begin{figure}
\centerline{\includegraphics{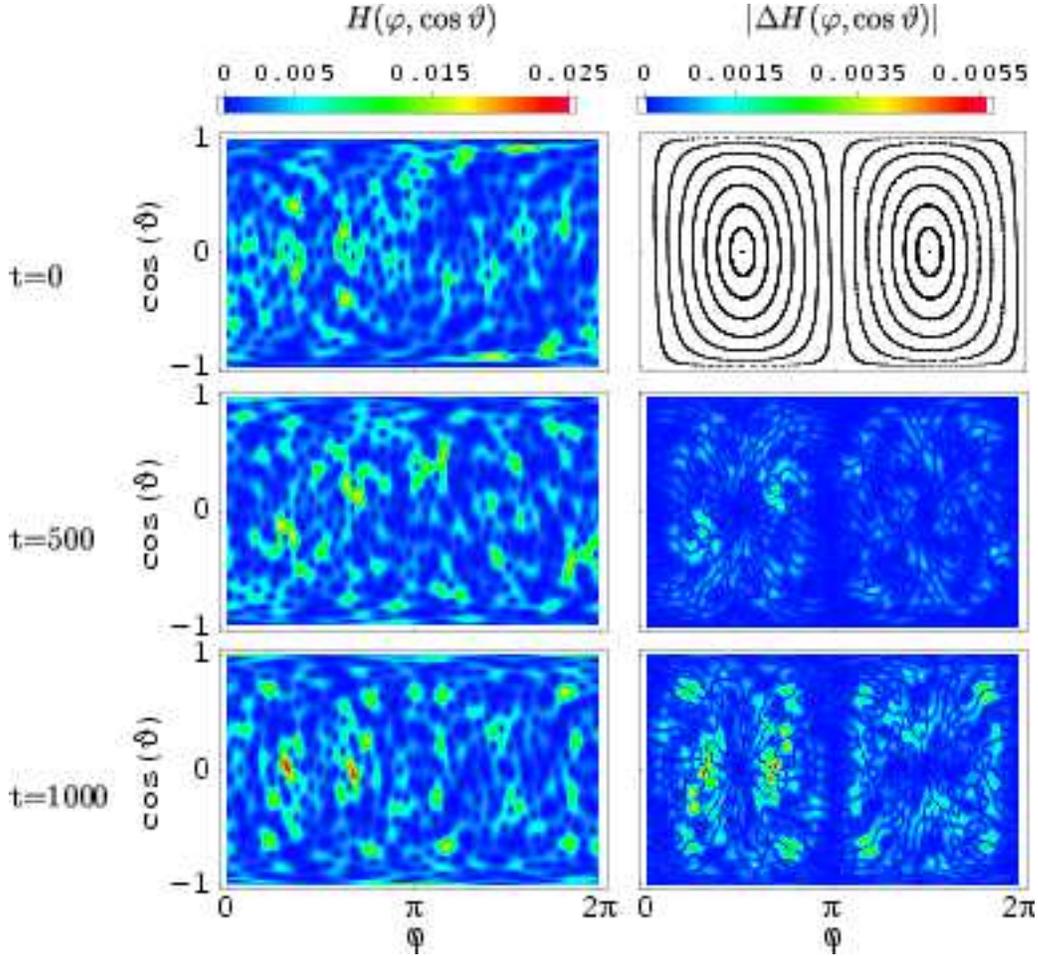}}
\caption{Husimi function $H(\varphi,\cos\vartheta)$ for a regular evolution $\alpha=0.1$, $\gamma=\pi/2$ and three different 
times $t=0,500$ and $1000$ (left column). Gaussian random initial state is used and $J=200$, $\delta=10^{-4}$. In the 
right column we show pictures of absolute difference between Husimi functions of the unperturbed and perturbed time 
evolution, $|H(\varphi,\cos\vartheta)-H_{\delta}(\varphi,\cos\vartheta)|$. Fidelity at the times shown is $|F(500)|=0.73$ and 
$|F(1000)|=0.25$. Top picture in the right column gives the classical phase space portrait.}
\label{fig:Hus1}
\end{figure}
\begin{figure}
\centerline{\includegraphics{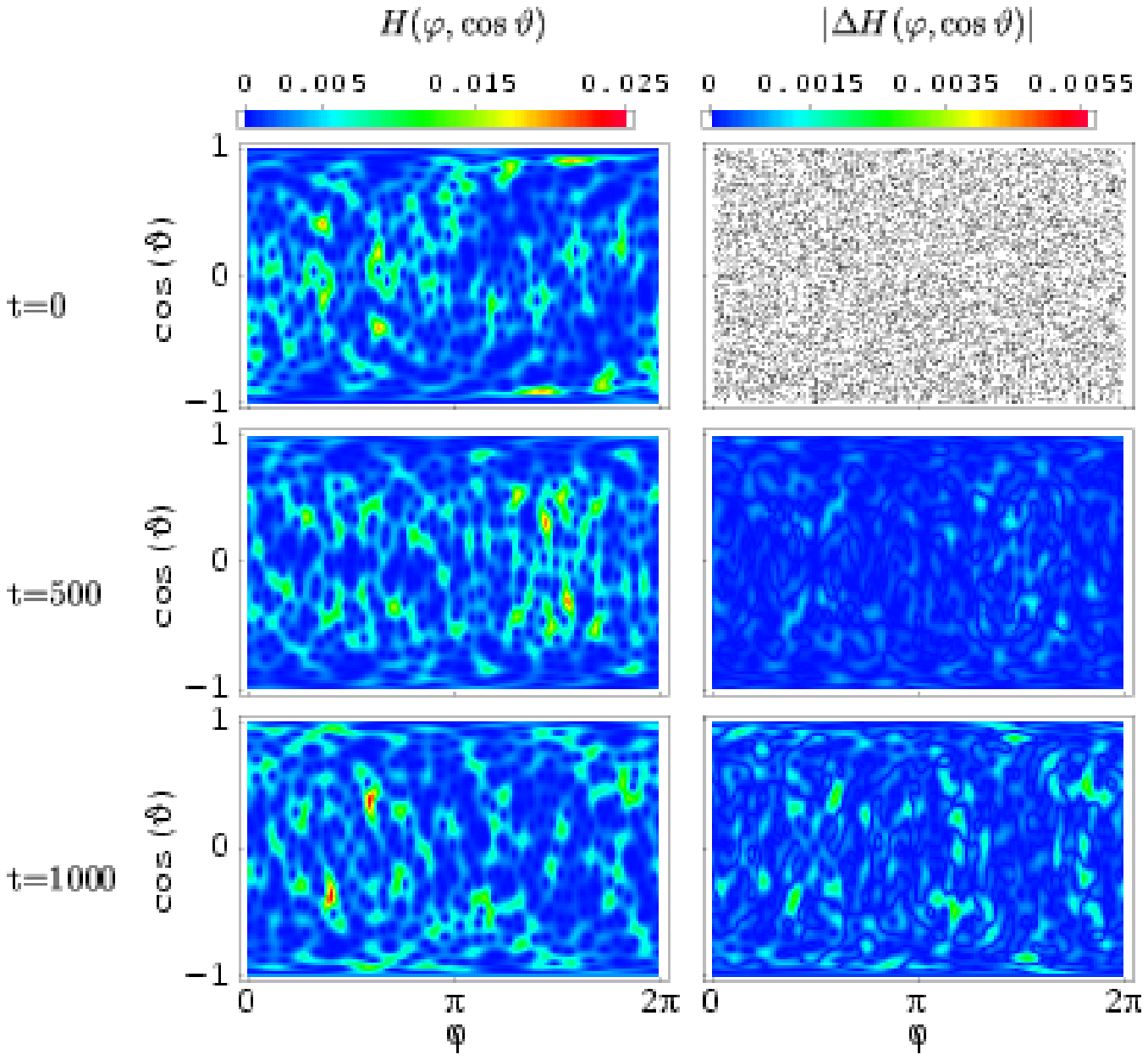}}
\caption{The same as in fig.~\ref{fig:Hus1} except for a chaotic (mixing) dynamics, $\alpha=30$.
Numerical values of the fidelity are now $|F(500)|=0.997$ and $|F(1000)|=0.988$.}
\label{fig:Hus2}
\end{figure}

Finally, we want to visualize the phenomenon of faster decay of fidelity in a regular, non-mixing regime $\alpha=0.1$ in 
comparison with the chaotic, mixing regime $\alpha=30$ by using the phase space representation of wave functions. 
A convenient and popular choice is a Husimi function $H(\varphi,\cos\vartheta)$ of a state $\ket{\psi}$ defined as
\begin{equation}
H(\varphi,\cos\vartheta) =|\braket{\vartheta,\varphi}{\psi}|^2.
\label{eq:Husimi}
\end{equation}
We have chosen a pure random initial state $\ket{\psi}$ and propagated it, in the first case with the propagator $U$ for 
$\alpha=0.1$, and in the second case for $\alpha=30$, as well as with the perturbed propagators $U_\delta$ in both cases. 
The state and all the other parameters (e.g. perturbation strength $\delta$) were identical in both cases. 
Then we compared the differences between the Husimi functions 
$\Delta H(\varphi,\cos\vartheta) = H(\varphi,\cos\vartheta) -H_\delta(\varphi,\cos\vartheta)$ 
of the perturbed and the unperturbed time evolution. 
The results for $\alpha=0.1$ are shown in fig.~\ref{fig:Hus1} and for $\alpha=30$ in fig.~\ref{fig:Hus2}. 
For chosen $\delta=10^{-4}$ the fidelity decay for the regular dynamics is much faster than for the chaotic one 
and this effect can also be observed in Husimi functions by comparing the right columns of both figures.
But one should stress that by considering the difference of Husimi functions $\Delta H = H - H_\delta$, 
we loose information about the relative phases of the perturbed and unperturbed wave functions 
in the coherent-state basis which are more important for the fidelity decay in the regular case
($\alpha=0.1$). However, results shown in figs.~\ref{fig:Hus1},\ref{fig:Hus2} 
suggest that not only the quantum phases but also
the amplitudes (in some classically motivated, e.g. coherent state basis) exhibit larger susceptibility 
to system perturbations for the regular (non-mixing) as compared  to the chaotic (mixing) dynamics.

\subsection{Pair of coupled kicked tops}

As we have already remarked at the end of sect.~\ref{sec:timepert}, for a one dimensional $d=1$ systems, 
the `surprising' behavior $\tau_{\rm ne-coh} < \tau_{\rm em}$ is for coherent initial states possible only around 
the border (\ref{eq:deltas}) $\delta_{\rm s}$ (unless $\sigma_{\rm cl}$ is very small like in the example of 
fig.~\ref{fig:parad1}) where the exponential decay in the mixing regime goes over to a gaussian due to finite size 
${\cal N}$. However, for more than one degree of freedom, $d>1$, such behavior is generally possible well above the
finite size -- perturbative border $\delta_{\rm s}$. In order to illustrate this phenomenon we will now briefly consider a 
numerical example of a pair of coupled kicked tops \cite{Miller} where $d=2$.
\begin{figure}
\centerline{\includegraphics{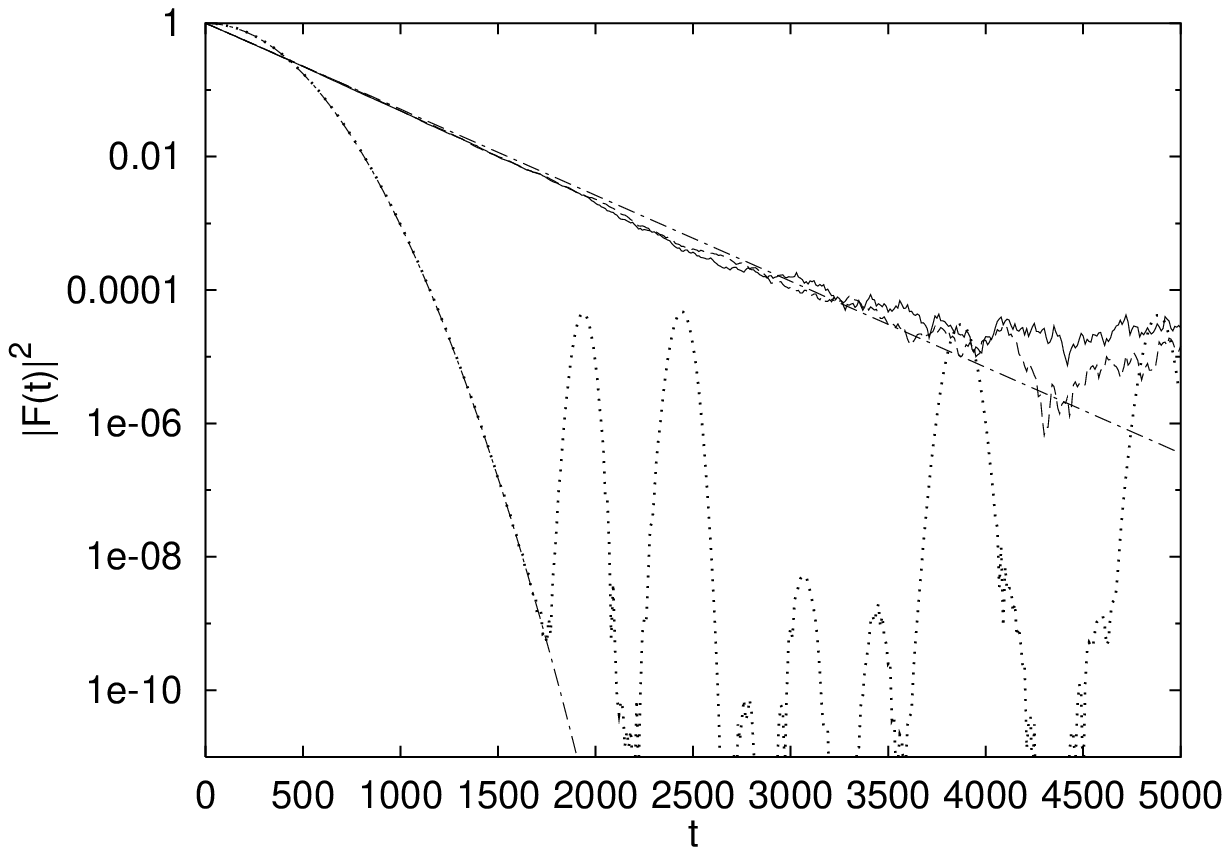}}
\caption{Fidelity for two coupled kicked tops, $\delta=8\cdot 10^{-4}$ and $J=200$. 
The upper curves are for $\epsilon=20$ (mixing regime), solid curve for a coherent initial state (\ref{eq:ket})
and dashed curve for a random initial state, and the lower -- dotted curve is for $\epsilon=1$ (non-mixing regime)
with a coherent initial state (\ref{eq:ket}).
The exponential and gaussian chain curves give, respectively, the expected theoretical decays 
(\ref{eq:Fmix}) and (\ref{eq:Fcoh}, here decay time is determined by the best fit).}
\label{fig:parad2}
\end{figure}   
\par
We consider coupled kicked tops with a unitary propagator 
(a simplified version of that of ref.~\cite{Miller})
\begin{equation}
U(\epsilon)=
e^{-i \frac{\pi}{2} J_{\rm 1y}} e^{-i \frac{\pi}{2} J_{\rm 2y}} e^{-i \epsilon J_{\rm 1z} J_{\rm 2z}/J}.
\label{eq:2KT}
\end{equation}
where $\ve{J_1}$ and $\ve{J_2}$ are two independent quantum angular momentum vectors.
Perturbed propagator is obtained by perturbing the parameter $\epsilon$, so that 
$U_\delta=U(\epsilon+\delta)$. The perturbation generator is therefore
\begin{equation}
A=\frac{1}{J^2} J_{\rm 1z} J_{\rm 2z},
\label{2KTA}
\end{equation}
with $\hbar=1/J$. A pair of coupled kicked tops possesses the same pair of discrete symmetries
as the single top for $\gamma=\pi/2$ \cite{Peres1}, namely 
$R_{\rm x}$ and $R_{\rm y}$, and in addition it is unvariant under the
permutation of the identical tops. However in our
simple-minded numerical experiment reported here we have used the propagator (\ref{eq:2KT}) over 
the full $(2J+1)^2$ dimensional Hilbert space thus making the appropriate average over all symmetry 
classes.
\par
The classical limit is obtained by $J\to \infty$ and writing the classical angular momentum vectors in terms of two unit 
vectors on the sphere $\ve{r}_{1,2}=\ve{J}_{1,2}/J$. In component notation we get the following equations of motion
\begin{eqnarray}
x_{1,2}'&=& z_{1,2}\\
y_{1,2}'&=& y_{1,2} \cos(\epsilon z_{2,1}) + x_{1,2} \sin(\epsilon z_{2,1}) \nonumber \\
z_{1,2}'&=&-x_{1,2} \cos(\epsilon x_{2,1})+y_{1,2} \sin(\epsilon z_{2,1}) \nonumber 
\label{eq:KT2class}
\end{eqnarray}
We have chosen two regimes, namely non-ergodic (KAM) regime for $\epsilon=1$ where the vast majority of classical 
orbits are stable, and the mixing regime for $\epsilon=20$ where no significant traces of stable classical orbits
were found and where the correlation sum was to a very good accuracy given by the first term only
\begin{equation}
\sigma \approx \frac{1}{2} C(0)=\frac{1}{2 J^4 {\cal N}} \tr{J_{\rm 1z}^2 J_{\rm 2z}^2}=\frac{1}{18}\left(1 + \frac{1}{J}\right)^2.
\label{eq:2KTsigma}
\end{equation}
Our motivation here was to compare the non-ergodic and mixing fidelity decays for the coherent initial state which is here the
dyadic product of SU(2) coherent states (\ref{eq:SU2coh})
\begin{equation}
\ket{\vartheta,\varphi}_{12}=\ket{\vartheta_2,\varphi_2}_2 \otimes \ket{\vartheta_1,\varphi_1}_1.
\label{eq:ket}
\end{equation} 
In fig.\ref{fig:parad2} we show the fidelity decay at $J=200$ and $\delta=8\cdot 10^{-4}$ 
in non-ergodic and mixing cases started from the
coherent state (\ref{eq:ket}) with $(\vartheta_1,\varphi_1)=(\vartheta_2,\varphi_2)=\pi(1/\sqrt{3},1/\sqrt{2})$. 
We find excellent agreement between the theortical predictions (\ref{eq:Fmix}) and (\ref{eq:Fcoh}) and the
numerics. Note that we are here already in the regime $\delta < \delta_c$ where the fidelity decay in ergodic-mixing
regime is slower $\tau_{\rm em} = 673$ than in non-ergodic regime $\tau_{\rm ne-coh}=379$.
In the ergodic-mixing regime ($\epsilon=20$) we show for comparison also the fidelity decay for a random initial state
which is (due to ergodicity) almost identical to the case of coherent initial state.
We note that overall fidelity decay here is similar as in a one-dimensional case,
however, the scaling of various time and perturbation scales on $\hbar=1/J$ is different as discussed in 
subsect.\ref{sec:timepert}.
We want to conclude this section with an interesting experimental observation, namely that in the non-ergodic regime
the revivals of fidelity (quantum reccurrences) beyond the time scale $t_*$ are much less pronounced in 
$2d$ than in $1d$, e.g. compare figs.\ref{fig:parad1} and \ref{fig:parad2}.

\section{Classical fidelity}

In this section we wish to contrast the surprising findings on the stability of quantum motion with
an application of the general idea (section 2) to a fully classical concept, namely to (unitary) 
Perron-Frobenius propagator of classical phase space densities. Let us consider a classical dynamical system being
given by an invertible measure (volume) preserving map $\ve{\phi} :\ve{x} \to \ve{\phi}(\ve{x})$ on a $D$ $(=2d)$ 
dimensional phase space ${\cal M}$. No additional properties of the map will be assumed except of being at least 
piece-wise differentiable so that the stability matrix
\begin{eqnarray}
\ma{M}(\ve{x}) &=& \frac{\partial\ve{\phi}(\ve{x})}{\partial\ve{x}}, \\
\ma{M}_t(\ve{x}) &=& \frac{\partial\ve{\phi}^{(t)}(\ve{x})}{\partial\ve{x}} 
= \ma{M}(\ve{x}_{t-1})\cdots\ma{M}(\ve{x}_1)\ma{M}(\ve{x}_0),\quad
\ve{x}_t = \ve{\phi}^{(t)}(\ve{x})
\end{eqnarray}
exists almost everywhere.
Unitary Perron-Frobenius propagator $U_{\rm cl}$ is defined on a phase space density from $\rho \in L^2({\cal M})$ as
\begin{equation}
(U_{\rm cl} \rho)(\ve{x}) = \rho(\ve{\phi}^{(-1)}(\ve{x})).
\end{equation}
The next step is to perturb the map in a most general way by composing it 
\begin{equation}
\ve{\phi}_\delta = \ve{\phi} \circ \ve{g}_\delta,
\end{equation}
with a near identity map $\ve{g}_\delta(\ve{x})$
which is generated by an arbitrary (smooth) vector field $\ve{a}(\ve{x})$ thru the $\delta-$flow
\begin{equation}
\frac{d\ve{g}_\delta}{d\delta} = \ve{a}(\ve{g}_\delta),\qquad
{\rm with\;initial\;conditions}\quad
\ve{g}_0(\ve{x}) = \ve{x}.
\end{equation}
Vector field $\ve{a}(\ve{x})$ should be divergence free
\begin{equation}
{\rm div\,}\ve{a}(\ve{x})\equiv 0
\end{equation}
to keep the perturbed map $\ve{\phi}_\delta$ volume preserving 
(or hamiltonian for symplectic maps).
Now we can write the perturbed unitary Perron-Frobenius propagator
\begin{equation}
(U_{{\rm cl},\delta}\rho)(\ve{x}) = \rho(\ve{g}^{(-1)}_\delta(\ve{\phi}^{(-1)}(\ve{x})))
= \left(U_{\rm cl}\exp\left(-\delta\,\ve{a}(\ve{x})\cdot\partial/\partial\ve{x}\right)\rho\right)(\ve{x})
\end{equation}
in the standardized general form (\ref{eq:U_d}) with the self-adjoint perturbation operator
$ A_{\rm cl} $
\begin{equation}
U_{{\rm cl},\delta} = U_{\rm cl}\exp(-i\delta\,A_{\rm cl}),
\quad A_{\rm cl} = -i\,\ve{a}\cdot\frac{\partial}{\partial\ve{x}}.
\label{eq:Acl}
\end{equation}

We define the 'classical fidelity' as the inner product between the phase space densities
propagated by two slightly different Perron-Frobenius propagators from the same fixed initial density
$\rho(\ve{x})$
\begin{equation}
F_{\rm cl}(t) = \int_{\cal M}\!\!d^D\!\ve{x}\;(U^t_{\rm cl,\delta}\rho)(\ve{x}) (U^t_{\rm cl}\rho)(\ve{x}) = 
\int_{\cal M}\!\!d^D\!\ve{x}\;\rho(\ve{\phi}_\delta^{(-t)}(\ve{x}))
                              \rho(\ve{\phi}^{(-t)}(\ve{x})).
\end{equation}
Note that $F_{\rm cl}(t)$ follows the corresponding quantum fidelity $F(t)$
(2) for very short times $t \ll t_{\rm E}$, since the latter can be
alternatively written as
$F(t) = \int d^D\ve{x} W(\ve{x},t)W_\delta(\ve{x},t)$ in terms of
Wigner functions $W(\ve{x},t)$ and $W_\delta(\ve{x},t)$ of the states
$U^t\ket{\psi}$ and $U^t_\delta\ket{\psi}$, respectively, corresponding
to the Liouville densities $U^t_{\rm cl}\rho$ and $U^t_{\rm
cl,\delta}\rho$ in (74) if $\rho(\ve{x}) = W(\ve{x},t=0)$.
$F_{\rm cl}(t)$ can be given a probabilistic interpretation, namely 
if $\rho(\ve{x})$ is a characteristic function on a set 
${\cal B}\subset{\cal M}$, $\rho(\ve{x})=1$ or $0$, for $\ve{x}\in{\cal M}$ or $\ve{x}\not\in{\cal M}$, respectively, 
with the volume ${\cal V}=\int_{\cal M}\!\!d^D\!\ve{x}\,\rho(\ve{x})$ then
${\cal V}^{-1}F_{\rm cl}(t)$ is a {\em probability} that a {\em perturbed} system with the initial condition chosen
at random from the set ${\cal B}$ is found, after time $t$, in the 
image of the same set ${\cal B}$ propagated for time $t$ with the {\em unperturbed} system.

In order to proceed formally along eqs. (\ref{eq:Fprod}-\ref{eq:F2nd}) we define and straightforwardly compute the 
`classical Heisenberg operators' 
\begin{equation}
A_{{\rm cl},t} := U_{\rm cl}^{-t} A_{\rm cl} U_{\rm cl}^t = 
-i \ve{a}_t(\ve{x})\cdot\frac{\partial}{\partial\ve{x}},\quad
\ve{a}_t(\ve{x}) :=
\ma{M}^{-1}_t(\ve{x})\ve{a}(\ve{\phi}^{(t)}(\ve{x})).
\end{equation}
Finally, plugging this into $\delta-$expansion (\ref{eq:Fsum}), observing that the first order
always vanishes since ${\rm div\,}\ve{a}_t \equiv 0$, and simplifying the second order by integrating by parts,
we obtain
\begin{equation}
F_{\rm cl}(t) = 1 - \frac{\delta^2}{2}\int_{\cal M}\!\!d^D\!\ve{x}\;
\left|\frac{\partial\rho(\ve{x})}{\partial\ve{x}}\cdot\sum_{t'=0}^{t-1}\ma{M}^{-1}_{t'}(\ve{x})\ve{a}(\ve{\phi}^{(-t')}(\ve{x}))\right|^2 + {\cal O}(\delta^3).
\label{eq:Fcl2}
\end{equation}
Note that now the RHS does not look like a correlation function at all! The reason for this lies in an
essentially different form of the generator of perturbation (\ref{eq:Acl}) which in classical case involves phase 
space derivatives invoking the stability matrix, while in quantum mechanics 
it is a simple quantization of the observable $\ve{a}(\ve{x})$.

Now, observe that the classical linear response formula (\ref{eq:Fcl2}) gives precisely the same result as one 
intuitively expects based on the stability of individual orbits. 
Indeed, in the case of chaotic dynamics, the integrand (for sufficiently long times $t'$) becomes dominated by the leading
eigenvalue of the stability matrix $\ma{M}_t$, namely the largest Lyapunov exponent $\lambda$, 
and (\ref{eq:Fcl2}) can be written as
\begin{equation}
F_{\rm cl-cha}(t) = 1 - {\rm const_1}\,\delta^2 \exp(2\lambda t)
+ {\cal O}(\delta^3)
\end{equation}
where the unspecified constant depends on the initial distribution $\rho(\ve{x})$.
This means the classical fidelity for chaotic dynamic decays on logarithhmically short time-scale
$\sim \log(1/\delta)/\lambda$.
On the other hand, in the case of regular dynamics, all eigenvalues of the stability matrices $\ma{M}_t$ have modulus $1$
so the integrand remains bounded by some constant and the fidelity can be estimated as
\begin{equation}
F_{\rm cl-reg}(t) = 1 - {\rm const_2}\,\delta^2 t^2 + {\cal O}(\delta^3)
\end{equation}
Note that the fidelity decay for regular dynamics has the same $\delta$ and $t$ dependence as in the
quantum chase (\ref{eq:Fr2}) with the decay time-scale $\sim \delta^{-1}$.

\section{Summary and discussion}

In this paper we have presented a simple theory on intrinsic decoherence of deterministic (isolated)
quantum dynamical systems due to small static perturbations of the evolution operator.
The central object of study is the fidelity $F(t)$ of quantum motion computed either with a single (pure) initial state
or averaged over an ensemble of initial states described by the statistical density matrix. 

The main result of the paper is a simple linear response (or Kubo-like) formula 
(\ref{eq:F2nd}) which relates the fidelity decay to the total sum (or integral) of two point time
autocorrelation function of the generator of perturbation.
In the limit of infinite Hilbert space dimension we found 
{\em exponential} fidelity decay on a time-scale $\tau_{\rm em} \propto (\hbar/\delta)^2$, for 
{\em quantum ergodic and mixing} systems, whereas for {\em non-ergodic} systems we have found much faster 
decay (in the sufficiently `quantum' regime where $\delta \ll \hbar$) on a time-scale 
$\tau_{\rm ne} \propto \hbar/\delta$ for random initial states or $\tau_{\rm ne-coh} \propto 
\hbar^{1/2}/\delta$ for coherent initial states (minimal uncertainty wavepackets)
where the fidelity $F(t)$ is given by a {\em Fourier transformation of the 
local density of states of the time averaged perturbation operator}.
A special emphasis was given to the semiclassical theory of fidelity of small but finite values of $\hbar$, 
where different regimes and the corresponding time and perturbation scales are carefully discussed, 
and where fidelity decay may asymptotically (as $\hbar\to 0$) be evaluated in terms of classical quantities only. 
Interestingly, finite size fluctuations of fidelity (for very long times at a finite Hilbert space dimension) 
have been shown to be given by the {\em inverse participation ratio} of the eigenstates of the perturbed 
evolution operator in the eigenbasis of the unperturbed propagator. 
The surprising aspects of our relations are mainly due to non-interchangability of the limits $\delta\to 0$ and 
$\hbar\to 0$ as the relevant decay time-scales are only functions of the ratio $\delta/\hbar$.
Therefore, a different and intuitively expected behaviour, namely faster fidelity decay for mixing than regular 
dynamics, is obtained in the `classical' regime where $\hbar \ll \delta$ 
(or making the limit $\hbar\to 0$ prior to $\delta\to 0$).
A similar, reassuring result has been found by applying our formalism to inspect the analogous 
classical fidelity for the unitary Perron-Frobenius evolution of volume (area) preserving maps: 
there the classical fidelity for regular dynamics has been found to decay on a time-scale $\propto \delta^{-2}$, 
while for a chaotic dynamics, the fidelity decay is governed by
a maximal Lyapunov exponent $\lambda$ on a short time-scale 
$\propto \ln(1/\delta)/\lambda$.

We note that our findings should be of primary importance for understanding the problem of stability of quantum
computation \cite{QC} where different regimes and time-scales \cite{DimaKR} 
should now be explained in terms of intrinsic dynamics of a particular quantum algorithm. 
Actually, our relation between fidelity and correlation decay may prove very useful in 
design of any new technology which is aimed at manipulating individual quantum states.
On the other hand, our results may also shed new light on the relation
between decoherence and dynamics \cite{Zurek}. In particular, due to quite special role of 
coherent initial states on the short Ehrenfest time-scale one may expect the results to be quite 
different for a random initial state and/or for longer time-scales (see \cite{Gorin}).

\section*{Acknowledgements}

We thank T. H. Seligman for very fruitful discussions and for many encouraging and useful remarks.
We also acknowledge discussions with G. Usaj, S. Tomsovic and G. Veble. The work has been supported
by the Ministry of Education, Science and Sport of Slovenia.

\end{document}